\documentclass[reprint,twocolumn,aps,prb,superscriptaddress]{revtex4-2}
\usepackage[T1]{fontenc}
\usepackage[utf8]{inputenc}
\usepackage{lmodern}
\usepackage{graphicx}
\usepackage{xcolor}
\usepackage{float}
\usepackage{siunitx}
\usepackage{hyperref}
\usepackage{braket}
\usepackage{amssymb,amsmath,amsthm,,amsfonts, array}
\usepackage{mdframed}
\usepackage{systeme,mathtools}
\usepackage{yfonts}
\usepackage{comment}
\usepackage{bm}

\DeclareUnicodeCharacter{0301}{-}

\usepackage{tikz}
\usepackage{pgfplots}
\pgfplotsset{compat=1.18}
\definecolor{tealc}{HTML}{0F6E56}
\definecolor{coralc}{HTML}{B0451D}
\definecolor{grayc}{HTML}{5F5E5A}

\newcommand{\bI}{\mathbf{I}}
\newcommand{\btheta}{\bm{\theta}}
\newcommand{\bomega}{\bm{\omega}}
\newcommand{\bk}{\mathbf{k}}
\newcommand{\R}{\mathbb{R}}
\newcommand{\Z}{\mathbb{Z}}
\newcommand{\T}{\mathbb{T}}
\newcommand{\eps}{\varepsilon}
\newcommand{\pb}[2]{\{#1,#2\}}
\newcommand{\dd}{\mathrm{d}}

\newcommand{\vk}{\bm{k}}
\theoremstyle{definition}
\newtheorem{remark}{Remark}
\usepackage{enumitem}

\newcommand{\vw}{\bm{\omega}}
\newcommand{\vth}{\bm{\theta}}
\newcommand{\vI}{\bm{I}}
\newcommand{\Wc}{W_c}
\DeclareMathOperator{\sech}{sech}

\usepackage{graphicx}
\usepackage{xcolor}

\newtheorem{theorem}{Theorem}
\newtheorem{proposition}{Proposition}
\newtheorem{lemma}{Lemma}
\newtheorem{corollary}{Corollary}

\newcommand{\pitch}{\pitchfork}
\newcommand{\gme}{g_{\mathrm{ME}}}
\newcommand{\Vm}{\mathcal V_m}

\usepackage{booktabs}
\usepackage[caption=false]{subfig}

\begin{document}

\renewcommand{\vec}[1]{\mathbf{#1}}
\newcommand{\ii}{\mathrm{i}}
\def\ya#1{{\color{orange}{#1}}}

\title{Separatrix splitting in a multiferroic neuromorphic
oscillator: toward Arnold diffusion}

\author{D. Maroulakos}
\affiliation{Doctoral School at the  University of Rzesz\'ow, 35-317 Rzesz\'ow, Poland}

\author{Z. Toklikishvili}
\affiliation{Faculty of Exact and Natural Sciences, Tbilisi State University, Chavchavadze av.3, 0128 Tbilisi, Georgia}

\author{A. Wal}
\affiliation{Institute of Physics, Faculty of Exact and Technical Sciences, University of Rzesz\'ow, 35-317 Rzesz\'ow, Poland}

\author{L. Chotorlishvili}
\affiliation{Department of Physics and Medical Engineering, Rzesz\'ow University of Technology, 35-959 Rzesz\'ow, Poland}

\date{\today}

\begin{abstract}

Prototype magnetic neuromorphic systems, in essence, are based either on ferromagnetic or antiferromagnetic materials as key elements for
conventional neuromorphic setups. In this work, we propose a novel neuromorphic system based on a ferromagnetic disk and a ferroelectric layer coupled via the magnetoelectric effect. We study a magnetoelectrically coupled ferromagnetic--ferroelectric oscillator as a candidate element for neuromorphic devices. A local reduction of conservative Landau--Lifshitz dynamics identifies a magnetic coordinate--momentum pair, which is coupled to an inertial
ferroelectric polarization mode. We derive the action--angle description of the resulting double-well--harmonic-oscillator Hamiltonian, obtain the quartic-well frequency map and Fourier coefficients in terms of elliptic functions, and establish monotone in-well twist. In the autonomous two-degree-of-freedom model, isoenergetic nondegeneracy supports KAM confinement on regular compact domains. Periodic electric forcing introduces a third effective frequency, although the harmonic magnetic sector leaves the unperturbed resonance loci parallel. An explicitly specified nonlinear magnetic extension produces transverse resonance intersections. We calculate leading resonant widths and local homoclinic splitting, and derive a damping-dependent splitting threshold for a stated scalar reduction. These results identify analytical structures relevant to transport and explain possible
polarization-based electrical readout. Establishing Arnold diffusion in the coupled model requires additional connecting-orbit analysis or controlled long-time evidence; neither a diffusion law nor Nekhoroshev stability of the driven model is inferred from the present finite-time calculations.
\end{abstract}

\maketitle

% ======================================================================
\section{Introduction}
% ======================================================================
Neuromorphic magnetic systems are a promising hardware platform for brain-inspired computing. In certain contexts, neuromorphic magnetic systems repeat the spiking behavior of biological neurons. Using intrinsic properties of electron spin and classical coarse-grained magnetization dynamics, neuromorphic magnetic systems emulate the energy-efficient, parallel processing of biological neural networks. The core building blocks of a prototype neuromorphic magnetic setup are spintronic devices that replace both synaptic and neuronal functions. Earlier neuromorphic proposals focus mainly on ferromagnetic (FM) materials, while more recent developments focus on antiferromagnetic (AFM) models. In particular, the recent work \cite{maroulakos2026} demonstrates a direct map between the AFM system and the celebrated FitzHugh--Nagumo model. In this work, we extend current FM and AFM neuromorphic magnetic systems and analyze a composite system of FM and ferroelectric (FE) layers coupled through magnetoelectric coupling. This work studies Arnold's diffusion in the FM-FE neuromorphic system. We describe the system of interest in detail in the model section. We begin our discussion with the cornerstone principles of Arnold's diffusion, which are considered throughout the paper.

A generic two-degree-of-freedom (2-DOF) Hamiltonian $H(p_1,p_2,q_1,q_2)$ possesses, in general, at least one conserved quantity, i.e., energy $E=\rm{const}$. The starting point is a splitting
\begin{eqnarray}\label{arnold1}
  H = H_0 + H_1 ,
\end{eqnarray}
in which $H_0$ is an integrable approximation (uncoupled normal modes, the quadratic part of a normal form, a separable truncation) and $H_1$ collects the non-integrable remainder.

For $H_0$ we invoke the Liouville--Arnold theorem: if there exist two integrals in involution, $H_0$ and $F$ with $\pb{H_0}{F}=0$, whose common compact regular level set is a two-torus, then there are canonical coordinates $(\bI,\btheta)\in\R^2\times\T^2$ with
\begin{eqnarray}\label{arnold2}
  I_j = \frac{1}{2\pi}\oint_{\gamma_j} p\,\dd q ,
  \qquad j = 1,2 ,
\end{eqnarray}
the integrals being taken over the two independent cycles $\gamma_j$ of the torus, and $H_0 = H_0(\bI)$ depends on the actions alone \cite{LichtenbergLieberman1992}. The conjugate angles evolve linearly, $\theta_j(t) = \theta_j(0) + \omega_j t$, with
\begin{eqnarray}\label{arnold3}
  \omega_j(\bI) = \frac{\partial H_0}{\partial I_j} .
\end{eqnarray}
Re-expressing the remainder in these coordinates and introducing a bookkeeping parameter $\eps$ for the coupling strength yields the canonical near-integrable form
\begin{eqnarray}\label{eq:nearint} 
&& H(\bI,\btheta) = H_0(\bI) + \eps\,H_1(\bI,\btheta),\\
&& H_1(\bI,\btheta) = \sum_{\bk\in\Z^2} h_{\bk}(\bI)\,e^{\,i\bk\cdot\btheta},\nonumber
\end{eqnarray}
with $H_1$ $2\pi$-periodic in each angle. Equation~\eqref{eq:nearint} is the object on which the entire analysis is built. We define the frequency map $\bomega(\bI) = \nabla_{\bI} H_0(\bI)$. The
\emph{resonance web} is
\begin{eqnarray}\label{arnold3}
  \mathcal{R}
  = \bigl\{\, \bI : \bk\cdot\bomega(\bI) = 0
    \text{ for some } \bk\in\Z^2\setminus\{0\} \,\bigr\}.
\end{eqnarray}
Here $\vk=(k_1,k_2)$ is an \emph{integer vector of Fourier harmonics}, one integer per angle. It enters because the perturbation $H_1(\vI,\vth)$ is $2\pi$-periodic in each angle and
hence admits the expansion $H_1=\sum_{\vk\in\Z^2}h_{\vk}(\vI)\,e^{\,i\vk\cdot\vth}$; only integer $\vk$ are compatible with that periodicity. The resonance condition
\begin{eqnarray}
\vk\cdot\vw(\vI)=k_1\omega_1(\vI)+k_2\omega_2(\vI)=0,
\end{eqnarray}
expresses a \emph{commensurability} (rational dependence) of the frequencies: the harmonic $e^{\,i\vk\cdot\vth}=e^{\,i\vk\cdot\vw\,t}$ then stops oscillating and drives the actions
secularly (the ``small-denominator'' resonance). The exclusion $\vk\neq 0$ removes the trivial mode: $\vk=0$ gives $0\cdot\vw=0$ for \emph{every} $\vI$, which would brand the whole
phase space resonant and carries no angle dependence to drive anything. So $\Z^2\setminus\{0\}$ is exactly the set of nontrivial driving harmonics whose vanishing combination of frequencies
defines a resonance line.

Two non-degeneracy conditions on $H_0$ are required. Kolmogorov non-degeneracy reads:
\begin{eqnarray}\label{arnold4}
\det\!\left(\frac{\partial^2 H_0}{\partial I_i\,\partial I_j}\right)= \det\!\left(\frac{\partial \omega_i}{\partial I_j}\right) \neq 0 ,
\end{eqnarray}
guarantees that $\bomega(\bI)$ genuinely sweeps a region and that Diophantine tori are abundant. The iso-energetic (Arnold) non-degeneracy, relevant for
diffusion \emph{on} a fixed energy shell, is
\begin{eqnarray}\label{arnold5}
  \det
  \begin{pmatrix}
    \dfrac{\partial^2 H_0}{\partial I_i\,\partial I_j}
      & \dfrac{\partial H_0}{\partial I_i} \\[12pt]
    \dfrac{\partial H_0}{\partial I_j} & 0
  \end{pmatrix}
  \neq 0 .
\end{eqnarray}
On a given resonance line $\bk\cdot\bomega = 0$, the harmonics $h_{\bk}(\bI)$ that do not vanish are the ones that drive the dynamics there.

For an \emph{autonomous} 2-DOF system Arnold diffusion does not occur, and this is a topological result rather than a matter of how small $\eps$ is. The energy surface
is three-dimensional, while surviving KAM tori are two-dimensional, hence of codimension one: they \emph{separate} the energy surface. Any trajectory is trapped in the gap between two adjacent tori, so the actions remain $\mathcal{O}(\sqrt{\eps})$-confined for all time \cite{LichtenbergLieberman1992}. What one \emph{can} obtain in 2 DOF is Chirikov resonance-overlap chaos---global stochasticity once neighbouring resonance layers overlap, governed by the overlap criterion---but this is not Arnold diffusion through a connected web \cite{Chirikov1979}. For physical application, e.g.,  quantum parametric nonlinear resonance see \cite{chotorlishvili2010quantum,PhysRevE.68.026216,PhysRevE.70.026219,PhysRevE.71.056211}. Arnold diffusion requires the invariant tori to have codimension at least two on the energy surface, so that their complement, the Arnold web, stays connected. This needs three frequencies. One therefore lifts the 2-DOF system in one of two ways. \paragraph{Time-periodic forcing.}
Take $H(\bI,\btheta,t) = H_0(\bI) + \eps H_1(\bI,\btheta,\nu t)$ and
autonomize by adjoining a pair $(I_3,\theta_3)$ with $\theta_3 = \nu t$:
\begin{eqnarray}\label{arnold6}
&&\widetilde H(I_1,I_2,I_3,\theta_1,\theta_2,\theta_3)= H_0(I_1,I_2) + \nu I_3\nonumber\\
&&+ \eps\,H_1(I_1,I_2,\theta_1,\theta_2,\theta_3).
\end{eqnarray}
The energy surface is now five-dimensional, the tori three-dimensional, of codimension two, and the web is connected.
\paragraph{A genuine third degree of freedom.}
Work directly with $H(I_1,I_2,I_3,\theta_1,\theta_2,\theta_3)$; the same counting applies.
\begin{remark}
If the physical model is truly autonomous and 2-DOF, the meaningful question is resonance overlap and transport, and ``Arnold diffusion'' proper enters only once a slow external drive or a third mode is included.
\end{remark}

\section{Model}

Multiferroics are materials in which multiple ferroic orders most notably ferroelectricity and magnetism coexist and couple either within in a  two distinct phases (\textbf{Type-I Multiferroics}: \textit{ferroelectricity and magnetism arise independently}), or single phase (\textbf{Type-II Multiferroics}: \textit{ferroelectricity is induced by magnetic order}). For both types of multiferroics, the key banchmark is the magneto-electric (ME) coupling, i.e., coupling between magnetization and ferroelectric polarization \cite{ramesh2007multiferroics,wang2009multiferroicity,fiebig2016evolution,cheong2007multiferroics}. The model proposed in this work is manly relevant to type-I Multiferroics and ME coupling plays a key role. 
\begin{figure}\label{fig:1}
\includegraphics[width=\columnwidth]{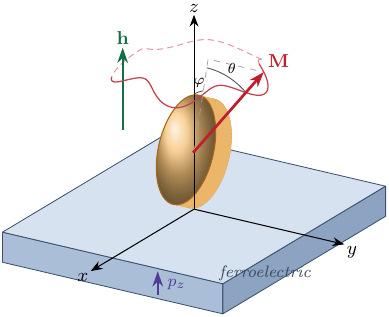}
\caption{The single-domain magnetic disk shaped nano-ellipsoid is deposited on a ferroelectric substrate. The uniaxial anisotropy has one easy axis along the z direction. Magnetoelectric coupling term $g_{\rm ME}M_zp_z$. Here $\bm M$ is the magnetization vector, $p_z$ is the ferroelectric polarization, $g_{\rm ME}$ is the coupling constant, and $\bm h$ is the magnetic field applied along the easy axis. Angles $\varphi$ and $\theta$ characterize 3D rotation of magnetization vector $\bm M$.} 
\end{figure}
We proceed with the structure of Hamiltonian and link this Hamiltonian to the prototype system discussed below.

In this work, we explore dynamics governed by the following two-dimensional $\bm q=(V, W)$, $\bm P=(p_V, p_W)$, time-dependent Hamiltonian $H(\bm P, \bm q, t)$: 
\begin{eqnarray}\label{eq: 1.model Hamiltonian}
&&H(\bm P, \bm q, t)=H_0(\bm P, \bm q)+\varepsilon H_1(\bm P, \bm q, t),\\
&&H_1(V, W, t)=\bigl(W-I_{\rm eff}(t)\bigr)V .\nonumber
\end{eqnarray}
Here, the unperturbed Hamiltonian has the form
\begin{eqnarray}\label{eq: 1.unperturbed Hamiltonian}
&&H_0(\bm P, \bm q)=\frac{p_V^2}{2m_V}+\frac{p_W^2}{2m_W}-\frac{V^2}{2}+\frac{V^4}{12}+\\
&&aW-\frac b2W^2,\nonumber
\end{eqnarray}
and $I_{\rm eff}(t)$ is the driving parameter. For further convenience, we introduce canonical action-angle variables $(I_v, \theta_V)$, $(I_W, \theta_W)$. To obtain three effective frequencies, we consider the protocol
\begin{eqnarray}
I_{\rm eff}(t)=I_{\rm dc}+F\cos\nu t,
\end{eqnarray}
Moreover, introduce the autonomous lift $\theta_3=\nu t$. Then Eq.(\ref{eq: 1.model Hamiltonian}) can be written in the form
\begin{eqnarray}\label{eq: 2.model Hamiltonian}
&&\widetilde H=H_0(I_V,I_W)+\nu I_3\\
&&+\varepsilon H_1(I_V,I_W,\theta_V,\theta_W,\theta_3).\nonumber
\end{eqnarray}
The model Eq.(\ref{eq: 1.model Hamiltonian}), i.e., Eq.(\ref{eq: 2.model Hamiltonian})  has straightforward physical interpretation: 
The magnetic subsystem precesses and couples to the ferroelectric polarization through magneto-electric coupling \cite{bibes2008towards,spaldin2005renaissance,PhysRevB.92.134424,PhysRevLett.96.067601,PhysRevB.91.041408,PhysRevB.94.104410,wanic2024magnetoelectric}. An external electric field drives the ferroelectric polarization.

\subsection{Magnetic coordinates and energy}

The system in question is shown in the pictorial plot Fig.1. We demonstrate that two transverse magnetization components form a coordinate--momentum pair. The quadratic energy of one component then becomes the kinetic term of the other component in canonical oscillator coordinates. Our construction uses conservative Landau--Lifshitz dynamics, and it does not require an additional microscopic spin-inertia term. The reduction below applies locally, for small transverse deviations and negligible damping over the time interval of interest.

We consider a uniformly magnetized particle with fixed saturation magnetization $M_s$ and reference direction along the $x$ axis. Let $\bm n$ be the unit vector $\bm n=(\cos\theta\sin\varphi,\, \sin\theta\sin\varphi,\, \cos\varphi)$. Then two independent components $n_y=Y=\sin\theta\sin\varphi$, $n_z=W=\cos\varphi$ are expressed as follows: 
\begin{eqnarray} \label{eq:coordinates}
&&\bm n=\frac{\bm M}{M_s}=\left(\sqrt{1-Y^2-W^2},Y,W\right),\\
&&W=\frac{M_z}{M_s},\quad  Y=\frac{M_y}{M_s},\nonumber
\end{eqnarray}
where $Y^2+W^2\ll1$. Thus, the variable $W$ is a normalized transverse \emph{component} of the magnetization, not its magnitude. We assume that the leading magnetic energy, including its coupling to the polarization coordinate $V$, has the form
\begin{eqnarray} \label{eq:energy} 
&& E=E_m+E_{\mathrm{ME}}=\frac{\kappa_y}{2}Y^2+U(W,V),\\
&& U(W,V)=\frac{\kappa_z}{2}W^2-h_zW+\gme VW.\nonumber
\end{eqnarray}
The coefficients $\kappa_y$ and $\kappa_z$ are the magnetic energy curvatures (magnetic anisotropy terms) with respect to the normalized transverse coordinates. In general, they may include contributions from crystalline anisotropy, demagnetizing fields, and an applied bias field. They can be controlled through the applied mechanical stress or doping magnetic impurities. A stable uncoupled reference state
requires positive curvatures. The term $-h_zW$ describes coupling to the external magnetic field, and $\gme VW$ describes magneto-electric coupling between the magnetization component $W$ and the ferroelectric polarization $V$. We treat these terms as perturbations because the force $h_z-\gme V$ must not tilt the magnetization outside the small-deviation regime. We use total energies for a magnetic volume $\Vm$. With $B_z$ the magnetic flux density in tesla, the Zeeman coefficient is
\begin{eqnarray} \label{eq:zeeman}
 h_z=M_s\Vm B_z.
\end{eqnarray}
If a magnetic field strength $H_z$ is used instead, the usual SI Zeeman convention gives $h_z=\mu_0M_s\Vm H_z$. These factors must be absorbed explicitly when using dimensionless coefficients.

\subsection{Reduction of Landau--Lifshitz dynamics}

The magnitude of the total spin angular momentum is
\begin{eqnarray} \label{eq:angularmomentum}
 \mathcal J=\frac{M_s\Vm}{|\gamma|},
\end{eqnarray}
where $|\gamma|$ is the gyromagnetic-ratio magnitude in units consistent with the magnetic flux density. We introduce a sign $s=\pm1$ to keep track
of the precession convention, and write the conservative equation as
\begin{eqnarray} \label{eq:LL}
\mathcal J\dot{\bm n}=s\,\bm n\times\frac{\partial E}{\partial\bm n}.
\end{eqnarray}
For example, the convention $\dot{\bm n}=-|\gamma|\bm n\times\bm B_{\mathrm{eff}}$ with $\bm B_{\mathrm{eff}}=-(M_s\Vm)^{-1}\partial E/\partial\bm n$ corresponds to $s=+1$. The result for the inertia will be independent of this sign. The conservative spin equation and its transverse-mode expansion are standard.
Expanding Eq.~\eqref{eq:LL} near $\bm n=(1,0,0)$ gives, to the retained order,
\begin{eqnarray} \label{eq:transverse}
\mathcal J\dot W=s\kappa_yY,\qquad \mathcal J\dot Y=-s\frac{\partial U}{\partial W}.
\end{eqnarray}
Define the momentum conjugate to $W$ by
\begin{eqnarray} \label{eq:momentum}
p_W=s\mathcal JY. 
\end{eqnarray}
We note that the dimension of $p_W$ and $\mathcal J$ is $[\rm Js]$, i.e., dimension of the action variable. For $s=-1$, this is the convention $p_W=-\mathcal JY$ used in the preceding explanation. Equation~\eqref{eq:transverse} becomes
\begin{eqnarray} \label{eq:canonical}
\dot W=\frac{\kappa_y}{\mathcal J^2}p_W,\qquad \dot p_W=-\frac{\partial U}{\partial W}.
\end{eqnarray}
These are Hamilton's equations for
\begin{eqnarray} \label{eq:effectiveH}
&&H_m+H_{\mathrm{ME}}=\frac{p_W^2}{2m_W}+U(W,V),\\
&& m_W=\frac{\mathcal J^2}{\kappa_y}.\nonumber
\end{eqnarray}
The dimension of the effective mass $m_W$ is $[\rm Js^2]$. The physical meaning of the kinetic term is therefore
\begin{eqnarray} \label{eq:meaning}
\frac{p_W^2}{2m_W}=\frac{\kappa_y}{2}Y^2. 
\end{eqnarray}
We note that the kinetic term is the magnetic energy stored in the second transverse component, expressed in terms of a canonical momentum. The variables $(W,p_W)$ describe one magnetic degree of freedom; the momentum encodes an existing spin component rather than adding an independent physical mode. Eliminating $Y$ yields
\begin{eqnarray} \label{eq:secondorder}
 m_W\ddot W+\kappa_zW-h_z+\gme V=0.
\end{eqnarray}
Equivalently, eliminating $p_W$ from $L=p_W\dot W-H_m-H_{\mathrm{ME}}$ gives
\begin{eqnarray}
 L_m=\frac{m_W}{2}\dot W^2-U(W,V),\qquad p_W=m_W\dot W.
\end{eqnarray}
Thus $m_W$ is an effective inertia, mass of the magnetic coordinate. For the dimensionless component $W$ and total-energy conventions, its units are $\mathrm{J\,s^2}$. Its value changes if the coordinate is rescaled. Thus $m_W$ is not a translational particle mass. The uncoupled small-amplitude precession frequency satisfies
\begin{eqnarray}
\omega_m^2=\frac{\kappa_z}{m_W}=\frac{\kappa_y\kappa_z}{\mathcal J^2}.
\end{eqnarray}

We note that the canonical oscillator description is a local expansion about a stable magnetic reference state. Its applicability requires
\begin{eqnarray} \label{corr:magneticdomain}
&&|W|\ll1,\qquad |Y|=|p_W|/J\ll1,\\
&&\kappa_y>0,\quad\kappa_z>0.\nonumber
\end{eqnarray}
The positive curvatures refer to the total magnetic energy, including demagnetizing, crystalline, and any bias-field contributions. An isolated
easy-axis term favoring $z$ is insufficient to stabilize an expansion about $x$; the assumed energy minimum must therefore be supported by the
complete magnetic energy. To the retained order, $p_W=sJY$ and $m_W=J^2/\kappa_y$ encode ordinary precessional dynamics, without an additional microscopic spin-inertia degree of freedom. When numerical coordinates of order unity are used, their relation to the physical transverse components must be specified. For example, setting $W_{\rm phys}=\eta_W w$ and $p_w=\eta_W p_{W,\rm phys}$ is canonical and gives
\begin{eqnarray} \label{corr:rescaledmagneticdomain}
\eta_W|w|\ll1,\qquad \frac{|p_w|}{\eta_W J}\ll1 .
\end{eqnarray}
The kinetic coefficient then becomes $m_w=m_W\eta_W^2$ before further time and energy rescalings. All coupling and curvature coefficients must be transformed with the same convention. Without this calibration, the numerical examples describe the dimensionless oscillator model, rather than a quantitatively specified magnetic device. The parameters $m_V$ and $m_W$ are effective inertial coefficients in this model.

\subsection{Coupling to the ferroelectric mode}

We choose the electric-field convention in which the field contribution to the polarization-mode energy is $-\mathcal E(t)P_{0z}V$. Here $\mathcal E$ is the external driving electric field coupled to dimensionless variable $V$. The conversion used in what follows includes the polarization-coordinate and volume normalizations: Polarization per unit volume  of the decoupled ferroelectric layer $p_{0z}$ has a dimension $[\rm Cm^{-2}]$ and net polarization of the whole layer $P_{0z}=\mathcal V_fp_{0z}$ has dimension $[\rm Cm]$, where $\mathcal V_f$ is the volume of the ferroelectric layer. The ME coupling term has the form $g_{\rm ME}VW$. Before transforming to the dimensionless case the reduced Hamiltonian reads
\begin{eqnarray} \label{corr:physicalH}
&&H_{\rm red}=\frac{p_V^2}{2m_V}+\alpha V^2+\beta V^4+\frac{p_W^2}{2m_W}+\frac{\kappa_z}{2}W^2\\
&&-h_zW+g_{\rm ME}VW-\mathcal E(t)P_{0z}V.\nonumber
\end{eqnarray}
The ferroelectric kinetic term has the form \cite{PhysRevB.92.134424}:
\begin{eqnarray}\label{1ferroelectric kinetic term}
K_F=\frac{1}{2}\rho\mathcal V_fp^2_{0z}\left(\frac{dV}{dt}\right)^2.
\end{eqnarray}

Here $\rho$ is an effective mass density for the polarization order parameter with the unit $[\rm Js^2mC^{-2}]$. For the transition to the dimensionless Hamiltonian we introduce the dimensionless time $\tau=\nu t$,  adopt the driving protocol $I_{\rm eff}(\tau)=\mathcal E_0+\mathcal E\cos\tau$, and rewrite the ferroelectric kinetic term in the form 
\begin{eqnarray}\label{2ferroelectric kinetic term}
\frac{p_V^2}{2m_V},\quad m_V=\rho\mathcal V_fp^2_{0z}\nu^2,\quad p_V=m_V\partial_\tau V. 
\end{eqnarray}  
It is easy to see that the dimension of the effective mass $m_V$ is Joule. Therefore, the transition to the dimensionless Hamiltonian is performed by operation 
$H(\tau)=H_{\rm red}/m_V$. Then explicitly we deduce:
\begin{eqnarray}\label{corr:H}
H(\tau)=H_0+\varepsilon V\bigl[W-I_{\rm dc}-F\cos(\tau)\bigr],
\end{eqnarray}
with unperturbed part
\begin{eqnarray}\label{corr:H0}
&&H_0=\frac{P_V^2}{2}+\frac{P_W^2}{2M_W}-\frac{V^2}{2}+\frac{V^4}{12}+\\
&&aW-\frac{b}{2}W^2. \nonumber
\end{eqnarray}
Here we introduce the following notations
\begin{eqnarray}\label{new notations}
&&a=-\frac{h_z}{m_V},\quad b=-\frac{\kappa_z}{m_V},\quad \alpha=-m_V/2=-6\beta, \nonumber\\
&&\varepsilon=\frac{g_{\rm ME}}{m_V},\quad I_{\rm dc}=\frac{P_{0z}\mathcal E_0}{g_{\rm ME}},\quad F=\frac{P_{0z}\mathcal E}{g_{\rm ME}},\\
&&M_W=\frac{\nu^2m_W}{m_V},\quad P_W=M_W\partial_\tau W,\quad \Omega_W=-b/M_W,\nonumber
\end{eqnarray}
and $P_V=p_V/m_V=\partial_\tau V$. Hamilton's equations give
\begin{eqnarray}\label{corr:Vequation}
&&\partial_{\tau\tau}V=V-\frac{V^3}{3}+\varepsilon\bigl[I_{\rm dc}+F\cos(\tau)-W\bigr],
\end{eqnarray}
\begin{eqnarray}\label{corr:Wequation}
&&M_W\partial_{\tau\tau}W=-a+bW-\varepsilon V.   
\end{eqnarray}
These equations define the conservative model used below. We note that for selected values of $\alpha$, the condition $\alpha=-\rho\mathcal V_fp^2_{0z}\nu^2$/2 connects the driving frequency $\nu$ with the material specific parameters, such as polarization per unit volume, effective mass density, and volume of the ferroelectric layer. The driving frequency can be tuned and controlled through those materiel specific parameters and geometry of the ferroelectric layer.  In what follows, the damping, where it is introduced, is specified separately at the level of the equations of motion. It is easy to see that Eq.(\ref{corr:H}), Eq.(\ref{corr:H0}) coincide with Eq.(\ref{eq: 1.model Hamiltonian}).

\section{Ferroelectric neuron concept}

The time-dependent polarization of a ferroelectric system, $\mathbf{P}(\mathbf{r},t)$, is experimentally accessible. The polarization dynamics can be converted into an electric potential or an electrical current and subsequently coupled to the membrane potential of an artificial neuron. This is a central concept of a ferro-neuromorphic system proposed in this work. The appropriate mapping depends on the geometry of the ferroelectric system and on the electrical boundary conditions.

\subsection{Polarization as a Source of Electric Field}

It is well-known that the spatially varying polarization produces a bound volume charge density
\begin{eqnarray} \label{eq:bound_charge}
\rho_{\mathrm{b}}(\mathbf{r},t)=-\boldsymbol{\nabla}\cdot\mathbf{P}(\mathbf{r},t),
\end{eqnarray}
while at the surface of the ferroelectric the corresponding bound surface charge density is
\begin{eqnarray}\label{eq:surface_charge}
\sigma_{\mathrm{b}}(\mathbf{r},t)=\mathbf{P}(\mathbf{r},t)\cdot\hat{\mathbf{n}},
\end{eqnarray}
where $\hat{\mathbf{n}}$ denotes the outward normal vector to the surface. The time-dependent polarization also generates a polarization current density,
\begin{eqnarray}\label{eq:polarization_current}
\mathbf{J}_{P}(\mathbf{r},t)=\frac{\partial\mathbf{P}(\mathbf{r},t)}{\partial t}.
\end{eqnarray}
For typical ferroelectric devices the electromagnetic wavelength is much
larger than the dimensions of the device. Therefore, the electrostatic or
quasi-electrostatic approximation can usually be employed.

\subsection{Electric Potential from the Polarization}

The electric displacement field is \cite{landau1984electrodynamics}:
\begin{eqnarray}\label{eq:D}
\mathbf{D}=\varepsilon_{0}\mathbf{E}+\mathbf{P},
\end{eqnarray}
and Gauss's law gives
\begin{eqnarray}\label{eq:gauss}
\boldsymbol{\nabla}\cdot\mathbf{D}=\rho_{\mathrm{f}},
\end{eqnarray}
where $\rho_{\mathrm{f}}$ is the free-charge density. Introducing the electrostatic potential
\begin{eqnarray}
\mathbf{E}=-\boldsymbol{\nabla}\phi,
\end{eqnarray}
one obtains
\begin{eqnarray}\label{eq:poisson_P}
\boldsymbol{\nabla}\cdot
\left[\varepsilon(\mathbf{r})\boldsymbol{\nabla}\phi(\mathbf{r},t)\right]
=\boldsymbol{\nabla}\cdot\mathbf{P}(\mathbf{r},t)-\rho_{\mathrm{f}}(\mathbf{r},t).
\end{eqnarray}
Equation~(\ref{eq:poisson_P}) provides a direct relation between the experimentally determined polarization distribution and the electric potential. For every experimentally measured time $t$, the polarization $\mathbf{P}(\mathbf{r},t)$ can be inserted into Eq.~(\ref{eq:poisson_P}), which can then be solved numerically with the appropriate boundary conditions. For a voltage-controlled electrode one imposes Dirichlet boundary conditions,
\begin{eqnarray}
    \phi=V_{\mathrm{ext}}(t),
\end{eqnarray}
whereas open-circuit or charge-controlled conditions can be formulated through the normal component of $\mathbf{D}$.

\subsection{Potential of a Spatially Resolved Polarization Distribution}

If the dielectric environment can approximately be considered homogeneous, the potential produced by the polarization distribution can also be expressed through the corresponding Green's function as \cite{jackson1999classical}:
\begin{eqnarray}\label{eq:green_potential}
&&\phi(\mathbf{r},t)=\frac{1}{4\pi\varepsilon}\Bigg[\int_{V}\frac{\rho_{\mathrm{b}}(\mathbf{r}',t)}{|\mathbf{r}-\mathbf{r}'|}\,d^{3}r'\\
&&+\int_{S}\frac{\sigma_{\mathrm{b}}(\mathbf{r}',t)}{|\mathbf{r}-\mathbf{r}'|}\,dS'\Bigg],
\end{eqnarray}
where
\begin{eqnarray}
\rho_{\mathrm{b}}=-\boldsymbol{\nabla}\cdot\mathbf{P},\qquad\sigma_{\mathrm{b}}
=\mathbf{P}\cdot\hat{\mathbf{n}}.
\end{eqnarray}
This formulation is particularly useful when the experiment provides spatially resolved polarization dynamics $\mathbf{P}(\mathbf{r},t)$.

\subsection{Parallel-Plate Ferroelectric Geometry}

We proceed with the discussion of the experimentally feasible setup of a prototype ferro-neuromorphic system.   A particularly simple relation can be obtained for a ferroelectric film of thickness $d$ placed between two parallel electrodes. Let the polarization be predominantly perpendicular to the film,
\begin{eqnarray}
    \mathbf{P}(t)=P_z(t)\hat{\mathbf{z}}.
\end{eqnarray}
The displacement field is then
\begin{eqnarray}\label{eq:D_parallel}
D_z=\varepsilon E_z+P_z.
\end{eqnarray}
Under ideal open-circuit conditions, and assuming vanishing free charge on the electrodes, one has approximately
\begin{eqnarray}
    D_z\simeq0.
\end{eqnarray}
Consequently,
\begin{eqnarray}\label{eq:E_P}
E_z(t)=-\frac{P_z(t)}{\varepsilon}.
\end{eqnarray}
The voltage between the electrodes is
\begin{eqnarray}
V_{\mathrm{FE}}(t)=-\int_{0}^{d}E_z(z,t)\,dz.
\end{eqnarray}
For a spatially uniform polarization this gives
\begin{eqnarray}\label{eq:V_FE}
V_{\mathrm{FE}}(t)=\frac{d}{\varepsilon}P_z(t).   
\end{eqnarray}
If the polarization is nonuniform along the thickness direction, the corresponding expression becomes
\begin{eqnarray}\label{eq:V_FE_nonuniform}
V_{\mathrm{FE}}(t)=\frac{1}{\varepsilon}\int_{0}^{d}P_z(z,t)\,dz.
\end{eqnarray}
Equation~(\ref{eq:V_FE}) therefore provides the simplest mapping from experimentally measured ferroelectric polarization dynamics to an
effective electrical voltage. It should be emphasized that this expression corresponds to ideal open-circuit conditions. In a real ferroelectric device, electrode screening, leakage currents, free carriers, and the impedance of the external circuit can modify this relation.

\subsection{Polarization Current}

An alternative and often more experimentally robust description follows from the polarization current. For a ferroelectric sample with electrode
area $A$, the polarization current is
\begin{eqnarray}\label{eq:I_P}
I_{P}(t)=A\frac{dP_z(t)}{dt}.
\end{eqnarray}
Therefore, experimentally measured polarization dynamics immediately determines the electrical current generated by the ferroelectric. If the ferroelectric is connected to an electrical load characterized by the impedance $Z(\omega)$, then in the frequency domain
\begin{eqnarray}\label{eq:load}
V_{\mathrm{FE}}(\omega)=Z(\omega)I_{P}(\omega).
\end{eqnarray}
This relation can be used to determine the voltage produced by the ferroelectric under realistic electrical loading conditions.

\subsection{Equivalent Circuit Description}

A convenient lumped-element representation of the ferroelectric consists of a polarization-current source in parallel with a dielectric
capacitance and a leakage resistance. The dielectric capacitance is approximately
\begin{eqnarray}\label{eq:C_FE}
C_{\mathrm{FE}}=\frac{\varepsilon A}{d}.
\end{eqnarray}
Introducing a leakage resistance $R_{\mathrm{FE}}$, the terminal voltage satisfies an equation of the form
\begin{eqnarray}\label{eq:FE_circuit}
C_{\mathrm{FE}}\frac{dV_{\mathrm{FE}}}{dt}+\frac{V_{\mathrm{FE}}}{R_{\mathrm{FE}}}=
I_{\mathrm{ext}}(t)-I_{P}(t),   
\end{eqnarray}
where the overall sign of $I_P$ depends on the chosen electrode and polarization conventions. Using Eq.~(\ref{eq:I_P}), this can equivalently be written as
\begin{eqnarray}\label{eq:FE_circuit_P}
C_{\mathrm{FE}}\frac{dV_{\mathrm{FE}}}{dt}+\frac{V_{\mathrm{FE}}}{R_{\mathrm{FE}}}=I_{\mathrm{ext}}(t)-A\frac{dP_z}{dt}.   
\end{eqnarray}
This equation provides a direct connection between experimentally observed polarization dynamics and the electrical dynamics of an artificial neuron.

\subsection{Coupling to an Artificial-Neuron Membrane Potential}

The membrane potential of an artificial neuron can generally be described by
\begin{eqnarray}\label{eq:membrane}
C_m\frac{dV_m}{dt}=-I_{\mathrm{ion}}(V_m,\ldots)+I_{\mathrm{in}}(t),
\end{eqnarray}
where $C_m$ denotes the effective membrane capacitance and
$I_{\mathrm{ion}}$ represents the internal nonlinear current responsible
for the neuron dynamics. There are two natural ways of coupling the ferroelectric dynamics to the
artificial neuron.

\subsubsection{Current-mediated coupling}

The ferroelectric polarization current can directly act as an input current,
\begin{eqnarray}\label{eq:neuron_current}
I_{\mathrm{in}}(t)=\eta I_P(t)=\eta A\frac{dP_z(t)}{dt},
\end{eqnarray}
where $\eta$ is an effective coupling coefficient. The membrane equation then becomes
\begin{eqnarray}\label{eq:membrane_FE_current}
C_m\frac{dV_m}{dt}=-I_{\mathrm{ion}}(V_m,\ldots)+\eta A\frac{dP_z(t)}{dt}.   
\end{eqnarray}
This formulation is particularly natural when polarization switching produces experimentally measurable current pulses.

\subsubsection{Voltage-mediated coupling}

Alternatively, under approximately open-circuit conditions, the ferroelectric-generated voltage can be introduced into the artificial neuron. Using Eq.~(\ref{eq:V_FE}),
\begin{eqnarray}
V_{\mathrm{FE}}(t)=\frac{d}{\varepsilon}P_z(t).
\end{eqnarray}
For example, if the ferroelectric voltage is coupled to the membrane through a resistance $R_c$, the corresponding input current can be written as
\begin{eqnarray}\label{eq:coupling_current}
I_{\mathrm{FE}\rightarrow m}=\frac{V_{\mathrm{FE}}(t)-V_m(t)}{R_c}.
\end{eqnarray}
Consequently,
\begin{eqnarray}\label{eq:coupled_membrane}
C_m\frac{dV_m}{dt}=-I_{\mathrm{ion}}(V_m,\ldots)+\frac{V_{\mathrm{FE}}(t)-V_m(t)}{R_c}.
\end{eqnarray}
After substitution of the experimentally determined polarization we obtain
\begin{eqnarray}\label{eq:coupled_membrane_P}
&&C_m\frac{dV_m}{dt}=-I_{\mathrm{ion}}(V_m,\ldots)+\\
&&\frac{1}{R_c}\left[\frac{d}{\varepsilon}P_z(t)-V_m(t)\right].\nonumber
\end{eqnarray}
Thus, the experimentally observed ferroelectric polarization dynamics can be incorporated directly into an artificial-neuron model.

\subsection{Connection with FitzHugh--Nagumo-Type Dynamics}

For a FitzHugh--Nagumo-type artificial neuron, we can write the following equation
\begin{eqnarray}\label{eq:FHN1}
C_m\dot V_m=f(V_m)-w+I_{\mathrm{FE}}(t),
\end{eqnarray}
where 
\begin{eqnarray}\label{eq:FHN2}
\tau_w\dot w=g(V_m,w),
\end{eqnarray}
and $w$ denotes the recovery variable. The ferroelectric dynamics can enter through either the polarization current 
\begin{eqnarray}\label{eq:FHN_FE_current}
I_{\mathrm{FE}}(t)=\eta A\dot P_z(t),
\end{eqnarray}
or through voltage coupling,
\begin{eqnarray}\label{eq:FHN_FE_voltage}
I_{\mathrm{FE}}(t)=\frac{1}{R_c}\left[\frac{d}{\varepsilon}P_z(t)-V_m(t)\right].
\end{eqnarray}
The first formulation implies that rapid changes or switching events of the ferroelectric polarization produce neuronal excitation. In contrast,
the second formulation makes the instantaneous polarization state itself act as the effective driving voltage. Therefore, these two couplings represent physically different neuromorphic mechanisms:
\begin{eqnarray}
P(t)\longrightarrow V_{\mathrm{FE}}(t)\longrightarrow V_m(t)
\end{eqnarray}
for voltage-mediated coupling, whereas
\begin{eqnarray}
\dot P(t)\longrightarrow I_P(t) \longrightarrow V_m(t)
\end{eqnarray}
describes current-mediated coupling. Which mechanism is appropriate depends on the experimental electrical boundary conditions and on how the ferroelectric element is connected to the artificial-neuron circuit.

% ======================================================================
\section{Mathematical program}
% ======================================================================

Here we formulate a mathematical program for ourself and we will try to implement at least part of these aims. 

\subsection{Geometric formulation: whiskered tori and transition chains}

Assume now $n=3$ effective frequencies. Near a \emph{simple} resonance $\bk\cdot\bomega(\bI^\ast)=0$ (rank one), a resonant canonical transformation to the slow angle $\psi = \bk\cdot\btheta$ followed by averaging over the fast angles produces a pendulum-like normal form,
\begin{eqnarray}
&&K = \tfrac12\,A(\bI)\,J_\psi^2 + \eps\,B(\bI)\cos\psi+\\
&&(\text{fast, transverse actions}),\nonumber
  \label{eq:pendulum}
\end{eqnarray}
so that along the resonance one obtains a one-parameter family of \emph{whiskered} (partially hyperbolic) invariant tori carrying stable and unstable manifolds $W^{s}$, $W^{u}$---Arnold's ``whiskers'' \cite{Arnold1964}. The transverse actions can drift only through these
hyperbolic structures.

The diffusion problem then becomes geometric.
\begin{enumerate}[label=\textbf{\arabic*.},leftmargin=2.2em]
\item \textbf{Splitting of separatrices.} Determine whether $W^{u}(T)$ and
  $W^{s}(T)$ split (intersect transversally) or coincide. The leading measure is
  the Melnikov function
  \begin{eqnarray}
    M(\theta_0)
    = \int_{-\infty}^{\infty}
      \pb{H_0}{H_1}\!\bigl(z_0(t);\theta_0\bigr)\,\dd t ,
  \end{eqnarray}
evaluated on an unperturbed separatrix orbit $z_0(t)$. Simple zeros of $M$ imply transversal homoclinic/heteroclinic intersections and hence a nonzero step. The scattering map across the whiskers quantifies each step
  \cite{DelshamsLlaveSeara2006}.
\item \textbf{Transition chain.} Construct a sequence of whiskered tori $T_1,T_2,\dots$ with $W^{u}(T_j)\pitchfork W^{s}(T_{j+1})$ for every $j$. A shadowing/obstruction argument then yields a genuine orbit following the chain and drifting by $\mathcal{O}(1)$ in action along the resonance line \cite{Arnold1964,DelshamsLlaveSeara2006}.
\item \textbf{Large-gap problem.} Secondary resonances inside the normally hyperbolic invariant manifold open gaps in the primary chain; these are bridged using secondary (lower-dimensional) tori and the scattering map \cite{DelshamsLlaveSeara2006}.
\end{enumerate}

A structural distinction worth fixing at the outset:
\begin{itemize}[leftmargin=1.6em]
\item \emph{A priori unstable}: $H_0$ already contains a hyperbolic (pendulum) factor, the splitting is $\mathcal{O}(\eps)$, and the Melnikov method applies directly.
\item \emph{A priori stable}: $H_0$ is fully elliptic and hyperbolicity is created only by the perturbation; the splitting is then exponentially small, $\sim e^{-c/\sqrt{\eps}}$. This is Arnold's original regime and demands exponential-asymptotics control of the Melnikov integral \cite{Arnold1964}.
\end{itemize}

\subsection{Quantitative time scale: the Nekhoroshev bound}

The rate is constrained by Nekhoroshev's theorem. If $H_0$ is steep (convexity suffices), then
\begin{eqnarray}\label{1Nekhoroshev bound}
  \bigl|\bI(t) - \bI(0)\bigr| \le C\,\eps^{\,b},
\end{eqnarray}
for
\begin{eqnarray}\label{2Nekhoroshev bound}
|t| \le T_0\,\exp\!\bigl(C\,\eps^{-a}\bigr),
\end{eqnarray}

with $a = b = 1/(2n)$ in the convex case \cite{Nekhoroshev1977}. The actions are thus effectively frozen over exponentially long times, and Arnold diffusion is precisely the slow $O(1)$ drift that eventually defeats this perpetual stability; the diffusion coefficient along the web inherits the same exponential smallness.

\subsection{Compact recipe}

\begin{enumerate}[label=\textbf{\arabic*.},leftmargin=2.2em]
\item Split $H = H_0 + \eps H_1$; build action--angle variables for $H_0$ and expand $H_1 = \sum_{\bk} h_{\bk}(\bI)\,e^{i\bk\cdot\btheta}$.
\item Compute $\bomega(\bI) = \nabla_{\bI} H_0$; check Kolmogorov and iso-energetic non-degeneracy.
\item If the system is autonomous 2-DOF, lift it---add periodic forcing or a third mode---to reach $n=3$ effective frequencies; otherwise only confinement, not diffusion, is possible.
\item Map the resonance web $\bk\cdot\bomega(\bI)=0$ and identify the dominant
  $h_{\bk}$ on each line.
\item Resonant normal form / averaging $\to$ pendulum $\to$ whiskered tori.
\item Compute the Melnikov/splitting; verify transversal intersections.
\item Assemble a transition chain (bridging large gaps with the scattering map /
  secondary tori) and shadow it; estimate the rate against the Nekhoroshev
  bound.
\end{enumerate}
Whether the model is a priori stable or unstable, and whether the coupling already supplies a hyperbolic direction, decides almost everything about the difficulty and should be settled first.

\section{Analytic solution}

We proceed with the announced program, implement several items and leave the rest for the future. For the Hamiltonian in Eq.~(\ref{eq: 1.model Hamiltonian}) we write the perturbation with an explicit
bookkeeping parameter,
\begin{eqnarray}
&&H(t)=H_0+\varepsilon H_1(t),\\
&&H_1(t)=\bigl(W-I_{\rm eff}(t)\bigr)V ,\nonumber
\end{eqnarray}
where $\varepsilon=g_{ME}/m_V$ is the small parameter of the theory and $I_{\rm eff}$ is the dimensionless driving field. To obtain three effective frequencies we take into account that
\begin{eqnarray}
I_{\rm eff}(t)=I_{\rm dc}+F\cos\nu t,
\end{eqnarray}
and introduce the autonomous lift $\theta_3=\nu t$
\begin{eqnarray}
&&\widetilde H=H_0(I_V,I_W)+\nu I_3\\
&&+\varepsilon H_1(I_V,I_W,\theta_V,\theta_W,\theta_3).\nonumber
\end{eqnarray}
The autonomous two-degree-of-freedom system should not be called an Arnold-diffusion model; in that case KAM curves separate the three-dimensional energy
surface and only resonance-overlap transport can occur.  The following construction is therefore applied only to the periodically driven, or otherwise three-frequency, model.
A first technical point is the sign of the quadratic $W$ sector.  The potential
$aW-bW^2/2$ is confining only if $b<0$.  Setting $\beta=-b>0$ and $W_c=a/b=-a/\beta$, one obtains
\begin{eqnarray}
&&\frac{p_W^2}{2m_W}+aW-\frac{b}{2}W^2=\frac{p_W^2}{2m_W}+\nonumber\\
&&\frac{\beta}{2}(W-W_c)^2-\frac{a^2}{2\beta},
\end{eqnarray}
so that
\begin{eqnarray}\label{the second frequency}
&&W(\theta_W)=W_c+\sqrt{\frac{2I_W}{m_W\Omega_W}}\cos\theta_W,\nonumber\\
&&\Omega_W=\sqrt{\frac{\beta}{m_W}}.
\end{eqnarray}
For $b\ge 0$ the written Hamiltonian has no global compact $W$ action; one must
instead use a local Birkhoff normal form or add a confining nonlinear correction. The $V$ subsystem has
\begin{eqnarray}
U_V(V)=-\frac{V^2}{2}+\frac{V^4}{12},
\qquad -\frac34<E_V<0
\end{eqnarray}
inside either well.  For the positive well the turning points are
\begin{eqnarray}
V_\pm(E_V)=\sqrt{3\pm\sqrt{9+12E_V}} .
\end{eqnarray}
The one-well action is
\begin{eqnarray}\label{The one-well action}
&&I_V(E_V)=\frac{1}{\pi}
\int_{V_-(E_V)}^{V_+(E_V)}
\sqrt{2M_V\left(E_V-U_V(V)\right)}\,dV,\nonumber\\
&&0<I_V<I_{\rm sep},
\end{eqnarray}
with the separatrix value for $M_W=1$
\begin{eqnarray}\label{with the separatrix value}
&&I_{\rm sep}=\frac{2\sqrt{M_V}}{\pi}=\frac{2}{\pi} .
\end{eqnarray}
The period can be evaluated in closed form.  Defining
$r_1^2=3-\sqrt{9+12E_V}$ and $r_2^2=3+\sqrt{9+12E_V}$,
\begin{eqnarray}\label{evaluated in closed form}
&&T_V(E_V)=\frac{\sqrt{24M_V}}{r_2}K\!\left(1-\frac{r_1^2}{r_2^2}\right),\\
&&\omega_V(I_V)=\frac{2\pi}{T_V(E_V(I_V))},\nonumber
\end{eqnarray}
where $K$ is the complete elliptic integral.  The three-frequency map is then
\begin{eqnarray}\label{three-frequency map}
\omega(I)=\bigl(\omega_V(I_V),\Omega_W,\nu\bigr).
\end{eqnarray}
We note that after autonomization, the angles are $\vth=(\theta_V,\theta_W,\theta_3)$ with $\theta_3=\nu t$, and the integer triple
$\vk=(k_V,k_W,k_3)$ is the harmonic index conjugate to them: $k_V$ counts oscillations in
$\theta_V$ (the anharmonic $V$-orbit), $k_W$ in $\theta_W$ (the $W$-oscillator), $k_3$ in the
drive phase. The perturbation is expanded by sampling the unperturbed angle tori,
\begin{eqnarray}\label{the unperturbed angle tori}
&&h_{\vk}(\vI)=\frac{1}{(2\pi)^3}\int_{\T^3}H_1(\vI,\vth)\,e^{-i\vk\cdot\vth}\,d^3\theta,\\
&&k=(k_V,k_W,k_3)\in\mathbb Z^3.\nonumber
\end{eqnarray}
Here $k_V\omega_V+k_W\Omega_W+k_3\nu=0$ is the corresponding three-frequency resonance. The coefficient $h_{\vk}(\vI)$ 
is the amplitude of the $\vk$-th driving term.
Numerically it means a three-dimensional FFT on the grid $(\theta_V,\theta_W,\theta_3)$. The resonance surface $\vk\cdot\vw(\vI)=0$ is \emph{homogeneous} in $\vk$: replacing $\vk$ by $m\vk$ ($m\in\Z\setminus\{0\}$) defines the \emph{same} surface. The condition $\gcd(k_V,k_W,k_3)=1$ keeps only \emph{primitive} integer vectors and so labels each distinct resonance exactly once,
without double counting its higher harmonics $2\vk,3\vk$. The primitive vector is also the \emph{lowest-order} representative, and the order $|\vk|_1=|k_V|+|k_W|+|k_3|$ controls strength: in a smooth perturbation the harmonic amplitude $|h_{\vk}|$ typically decays with $|\vk|_1$. The primitive resonances are the roots of
\begin{eqnarray}\label{resonances are the roots of}
&&k_V\omega_V(I_V)+k_W\Omega_W+k_3\nu=0,\\
&&\gcd(k_V,k_W,k_3)=1 .\nonumber
\end{eqnarray}

For each resonance with $h_k\ne 0$ we form the rank-one normal form
\begin{eqnarray}\label{ rank-one normal form}
&&K_{\rm res}=\tfrac12A_{\bm k}J_\psi^2+2\eps|h_{\bm k}(I^*)|\cos\psi+\dots,\nonumber\\
&& A_k=k^T D_I^2H_0(I_*) k .
\end{eqnarray}
In the minimal quadratic-$W$ model, $A_{\bm k}=k_V^2\,d\omega_V/dI_V$. Every driving resonance has $k_V\neq0$: by Eq.~(98) the only nonzero harmonics with $k_V=0$ are $(0,\pm1,0)$ and $(0,0,\pm1)$, whose resonance conditions $\Omega_W=0$ and $\nu=0$ cannot be met. Hence $A_{\bm k}\neq0$ on every resonance, and the degeneracy $\det D_I^2H_0=0$ manifests itself not in a vanishing pendulum curvature but in the geometry of the web (parallel resonance lines, no junctions) and in the failure of Kolmogorov non-degeneracy and steepness. The corresponding full pendulum width estimate is
\begin{eqnarray}\label{pendulum width estimate is}
&&\Delta J_\psi=4\sqrt{\frac{2\eps|h_{\bm k}|}{|A_{\bm k}|}}=\frac{4}{|k_V|}\sqrt{\frac{2\eps|h_{\bm k}|}{|\omega_V'(I_V)|}},\\
&&\Delta I_V=|k_V|\,\Delta J_\psi .\nonumber
\end{eqnarray}
The a-priori-unstable channel associated with the $V$ saddle can be checked by a Melnikov calculation.  At $E_V=0$ the homoclinic orbit reads
\begin{eqnarray}
V_s(t)=s\sqrt6\,\operatorname{sech}\left(\frac{t}{\sqrt{M_V}}\right),
\qquad s=\pm1 .
\end{eqnarray}
Writing
$W(t)=W_c+A_W\cos(\Omega_W t+\phi_W)$ with
$A_W=\sqrt{2I_W/(m_W\Omega_W)}$, the phase-dependent Melnikov potential reads
\begin{eqnarray}\label{Melnikov potential reads}
&&L(\phi_W,\phi_3)=-A_W C(\Omega_W)\cos\phi_W+\\
&&F C(\nu)\cos\phi_3,\nonumber
\end{eqnarray}
up to the phase-independent term, where
\begin{eqnarray}\label{the phase-independent term}
C(\lambda)=s\sqrt6\,\frac{\pi\sqrt{M_V}}{\cosh\left(\pi\lambda\sqrt{M_V}/2\right)} .
\end{eqnarray}
We note that Eq.(\ref{the phase-independent term}) corresponds to the conservative, per-harmonic splitting amplitude. It is the amplitude of the Melnikov potential Eq.(\ref{Melnikov potential reads}), whose gradient gives the scattering kick $\delta I\propto\varepsilon\nabla L$: Eq.(\ref{the phase-independent term}). It carries \emph{no} damping, depends on frequency only, and decays. Crucially, for any finite $\lambda$ it is nonzero, so $L$ always has nondegenerate critical points and $W^u\pitch W^s$ transversally: there is \emph{no threshold}. This is the building block of the transition chain for Arnold diffusion in the conservative lift $H_0+\varepsilon H_1$.

The scattering-map kick is proportional to $\varepsilon\nabla L$.  Therefore a nonzero $A_WC(\Omega_W)$ or $FC(\nu)$ gives the leading separatrix splitting; simple nondegenerate critical points of $L$ give transverse intersections.

Finally, the code must report the non-degeneracy status rather than assume it. For the exact quadratic-$W$ and periodic-lift model,
\begin{eqnarray}\label{and periodic-lift model}
\det D_I^2H_0=0,\qquad\det D_I^2\widetilde H_0=0,
\end{eqnarray}
because the $W$ oscillator is isochronous and $\widetilde H_0$ is linear in $I_3$.  Thus the numerical program can map resonances, compute harmonics, normal-form widths, and Melnikov splitting, but strict theorem-level Kolmogorov/Nekhoroshev conclusions require restoring twist/steepness, for example by adding nonlinear $W$ or third-mode terms.

Taking into account Eq.(\ref{eq: 1.model Hamiltonian}), we write the perturbation as a product of single-angle functions,
\begin{eqnarray}\label{single-angle functions}
H_1=V(\theta_V)\,\big[\,W(\theta_W)-I_\mathrm{dc}-F\cos\theta_3\,\big].
\end{eqnarray}
Two of the three factors are \emph{pure cosines}:
\begin{eqnarray}\label{1the three factors are}
&&W(\theta_W)=\Wc+R_W\cos\theta_W,\\
&&R_W=\sqrt{\tfrac{2I_W}{m_W\Omega_W}},\nonumber
\end{eqnarray}
\begin{eqnarray}\label{2the three factors are}
&&I_\mathrm{eff}=I_\mathrm{dc}+F\cos\theta_3,
\end{eqnarray}
so they contribute only the harmonics $k_W\in\{0,\pm1\}$ and $k_3\in\{0,\pm1\}$. The triple
integral factorizes,
\begin{eqnarray}\label{3the three factors are}
&&h_{\vk}=v_{k_V}\cdot g_{k_W}\cdot f_{k_3},
\end{eqnarray}
where $v_{k_V}$ is the Fourier coefficient of the anharmonic $V$-orbit and the other two factors
are read off term by term. Collecting the terms:
\begin{eqnarray}\label{4the three factors are}
&&h_{(k_V,0,0)}=(\Wc-I_\mathrm{dc})\,v_{k_V},\nonumber\\
&&h_{(k_V,\pm1,0)}=\tfrac{1}{2}R_W\,v_{k_V},\\
&&h_{(k_V,0,\pm1)}=-\tfrac{1}{2}F\,v_{k_V},\nonumber
\end{eqnarray}
and $h_{\vk}=0$ whenever $|k_W|\ge 2$, or $|k_3|\ge 2$, or both $|k_W|=|k_3|=1$. The only
non-elementary part is
\begin{eqnarray}\label{The only non-elementary part}
v_{k_V}=\frac{1}{2\pi}\int_0^{2\pi}V(\theta_V)\,e^{-ik_V\theta_V}\,d\theta_V,
\end{eqnarray}
the harmonic content of the quartic-well orbit. That orbit is a Jacobi elliptic function. For the double-well $U_V=-V^2/2+V^4/12$ we will follow \cite{wanic2024magnetoelectric}:
The critical points are as follows:
\begin{eqnarray}
\begin{array}{c|c|}
\text{Quantity} &  H_{\mathrm{V}}\\
\hline
\text{centers} & (\pm\sqrt{3},0)\\[2mm]
\text{saddle} &  (0,0)\\[2mm]
\text{center energy} & -\dfrac34\\[2mm]
\text{separatrix energy} &  0\\[2mm]
\text{turning points on separatrix} &  V=\pm\sqrt{6}
\end{array}\nonumber
\end{eqnarray}
The homoclinic orbit
\begin{eqnarray}\label{1homoclinic orbit}
&&V^\pm(t)=\pm\sqrt{6}\,\operatorname{sech}(t-t_0),\\
&&\dot{V}^\pm(t)=\mp\sqrt{6}\,\operatorname{sech}(t-t_0)\tanh(t-t_0).\nonumber
\end{eqnarray}
The periodic solutions inside either well in terms of the elliptic Jacobi functions read: 
\begin{eqnarray}\label{2homoclinic orbit}
&&\dot{V}_<^\pm(t)=\mp\mathcal{B}(k)\operatorname{sn}\!\left(\frac{t-t_0}{\sqrt{2-k^2}},k\right)\operatorname{cn}\!\left(
\frac{t-t_0}{\sqrt{2-k^2}},k\right).\nonumber\\
&&V_<^\pm(t)=\pm\frac{\sqrt{6}}{\sqrt{2-k^2}}\,\operatorname{dn}\!\left(\frac{t-t_0}{\sqrt{2-k^2}},k\right),
\end{eqnarray}
where $\mathcal{B}(k)=\frac{\sqrt{6}\,k^2}{2-k^2}$.
The corresponding energy 
\begin{eqnarray}\label{3homoclinic orbit}
&& E(k)=3\,\frac{k^2-1}{(2-k^2)^2}<0.
\end{eqnarray}
The solution above the separatrix $E>0$: 
%\begin{widetext}
\begin{eqnarray}\label{4homoclinic orbit}
&&\dot{V}_>(t)=-\mathcal{F}_k\operatorname{sn}\!\left(\frac{t-t_0}{\sqrt{2k^2-1}},k\right)\operatorname{dn}\!\left(\frac{t-t_0}{\sqrt{2k^2-1}},k\right).\nonumber\\
&&V_>(t)=\frac{\sqrt{6}\,k}{\sqrt{2k^2-1}}\,\operatorname{cn}\!\left(\frac{t-t_0}{\sqrt{2k^2-1}},k\right)
\end{eqnarray}
%\end{widetext}
where $\mathcal{F}_k=\frac{\sqrt{6}\,k}{2k^2-1}$. The corresponding energy 
\begin{eqnarray}\label{3homoclinic orbit}
&& E(k)=3\,\frac{k^2(1-k^2)}{(2k^2-1)^2}.
\end{eqnarray}
The resonant normal form is
\begin{eqnarray}\label{4homoclinic orbit}
&&K_\mathrm{res}=\tfrac12 A_{\vk}J_\psi^2+2\varepsilon|h_{\vk}(\vI^\ast)|\cos\psi+(\text{fast terms}),\nonumber\\
&& A_{\vk}=\vk^{\mathsf T}D_I^2H_0(\vI^\ast)\,\vk.
\end{eqnarray}

Both ingredients are now in hand. The amplitude $|h_{\vk}|$ comes from Q4. The curvature $A_{\vk}$ needs the action Hessian, and here the model's degeneracy makes it almost trivial:
\begin{eqnarray}\label{5homoclinic orbit}
&&D_I^2H_0=
\begin{pmatrix}
\omega_V'(I_V) & 0 & 0\\[2pt]
0 & 0 & 0\\[2pt]
0 & 0 & 0
\end{pmatrix},\\
&&\omega_V'\equiv\frac{d\omega_V}{dI_V},\nonumber
\end{eqnarray}
because $\Omega_W$ is action-independent (the quadratic $W$-well is isochronous) and
$\tilde H_0$ is linear in $I_3$. Hence
\begin{eqnarray}\label{6homoclinic orbit}
A_{\vk}=k_V^2\,\omega_V'(I_V),
\end{eqnarray}
with the twist obtained from the elliptic period $T_V(E_V)$ by the action-angle
identities $dE_V/dI_V=\omega_V$ and $\omega_V=2\pi/T_V$:
\begin{eqnarray}
\omega_V'(I_V)=\frac{d\omega_V}{dI_V}
=\frac{d\omega_V}{dE_V}\frac{dE_V}{dI_V}
=-\frac{4\pi^2}{T_V^3}\,\frac{dT_V}{dE_V}.
\end{eqnarray}
The pendulum width then follows in closed form,
\begin{eqnarray}\label{7homoclinic orbit}
\Delta J_\psi=4\sqrt{\frac{2\varepsilon|h_{\vk}|}{|A_{\vk}|}}
=\frac{4}{|k_V|}\sqrt{\frac{2\varepsilon|h_{\vk}|}{|\omega_V'(I_V)|}} .
\end{eqnarray}
Following \cite{wiggins2003introduction}, with the damping $-\delta V'$ added to Eq.~(\ref{corr:H0}) ($\delta=O(\eps)$), the Melnikov's function: 
\begin{eqnarray}\label{The Melnikov function1}
&&M_\pm(t_0)=-4\delta\pm\sqrt6\,\pi\,\eps F\,\nu\,\sech\!\Big(\frac{\pi\nu}{2}\Big)\sin(\nu t_0),\nonumber\\
&&\sqrt6\,\pi\,\eps F\,\nu\,\sech(\pi\nu/2)>4\delta,\\
&&F_{\rm crit}(\nu)=\frac{4\delta}{\sqrt6\,\pi\,\eps}\,\frac{\cosh(\pi\nu/2)}{\nu},\\
&&F_{\rm crit}(\nu^*)\simeq1.232\,\frac{\delta}{\eps},\\
&&F_{\rm crit}\sim\frac{4\delta}{\sqrt6\pi\eps\nu}\ (\nu\to0^+),\\
&&F_{\rm crit}\sim\frac{2\delta\,e^{\pi\nu/2}}{\sqrt6\pi\eps\nu}\ (\nu\to\infty).
\end{eqnarray}
where $\delta$ is the damping constant. This phenomenological damping constant corresponds to the FE system and it was neglected in Eq.(\ref{eq: 1.model Hamiltonian}).
We note that $M_\pm(t_0)$ is the dissipative, time-domain Melnikov function. It has two pieces of \emph{different origin}: a constant offset $-4\delta$ from the
\emph{dissipation} (the phase-space area $\delta\!\int\dot V_s^2\,dt$ absorbed per homoclinic loop), and an oscillatory \emph{forcing} term. Note the extra factor $\nu$ relative to $C$:
$M$ is the time-domain Melnikov function $M=\int\{H_0,H_1\}\,dt$, which is the $t_0$-derivative of the potential, and $\partial_{t_0}\cos\nu t_0$ brings down $\nu$. Hence the forcing amplitude of Eq.(\ref{The Melnikov function1}) is exactly $\nu\,|C(\nu)|\,(F/\varepsilon)$ with $m_V=1$; the two expressions share the \emph{same} conservative content in the oscillatory term, related by $M_{\rm forcing}=\partial_{t_0}L$.

We note that the minimal model studied here has a degenerate frequency map whose resonance loci are parallel in the $(I_V,I_W)$ plane. A nonlinear recovery-sector frequency creates transverse resonance intersections and changes this geometry. These observations identify candidate transport structures but establish neither the absence nor the presence of Arnold diffusion. A demonstration of transport requires additional analysis of the relevant invariant manifolds, scattering dynamics, and connecting trajectories, or numerical evidence of substantial action drift with controlled integration errors.

\section{Results and analysis}\label{sec:results}

%----------------------------------------------------------------------
\subsection{Resonance web, layers, and overlap}

The web lives in action space. For each primitive $\bm k=(k_V,k_W,k_3)$ we draw the level set
\begin{equation}
k_V\,\omega_V(I_V)+k_W\,\Omega_W+k_3\,\nu=0 .
\label{eq:webline}
\end{equation}
In the minimal (degenerate) model only $\omega_V$ depends on an action [Eq.~(86)], while $\Omega_W$ [Eq.~(81)]
and the drive frequency $\nu$ are constants. Each resonance therefore fixes a value of $I_V$,
\begin{equation}
\omega_V(I_V^*)=-\frac{k_W\Omega_W+k_3\nu}{k_V},
\label{eq:Istar}
\end{equation}
and in the $(I_V,I_W)$ plane the web is a family of \emph{vertical lines} $I_V=I_V^*(\bm k)$, independent of $I_W$.
Once a nonlinear magnetic sector restores $\Omega_W=\Omega_W(I_W)$, the same conditions become genuine curves in
$(I_V,I_W)$ and a two-dimensional web with junctions appears.
Attaching to each line a band of full width $\Delta I_V=|k_V|\Delta J_\psi$ [Eq.~(91)] turns the web of zero-width
lines into \emph{layers} of finite-width islands.

Two distinct questions must not be conflated. \emph{Crossings} of two resonance lines are junctions (double
resonances). There several harmonics act at once, and transition chains can switch from one line to another; these
are the organizing centers of diffusion. The onset of \emph{global} stochasticity, by contrast, is governed by the
Chirikov overlap of neighboring layers~[3--7]. When the combined widths of adjacent resonances exceed their spacing,
\begin{equation}
S=\frac{\tfrac12(\Delta J_{\bm k}+\Delta J_{\bm k'})}{|J_{\bm k}-J_{\bm k'}|}\gtrsim1,
\label{eq:overlapS}
\end{equation}
the KAM curves between them are destroyed and the motion becomes globally chaotic. Crossings thus mark \emph{where}
Arnold routing can happen, and overlap marks \emph{whether} the chaos is global. Arnold diffusion proper belongs to
the opposite regime, $S\ll1$: isolated thin layers, with motion creeping \emph{along} a connected web rather than across
overlapped layers.

\begin{table*}[tbh]
\label{tab:resonances}
\begin{ruledtabular}
\begin{tabular}{lcccccccc}
$\bm k$ & $|\bm k|_1$ & $\omega_V^*$ & $I_V^*$ & $E_V^*$ & $|v_{k_V}|$ & $|h_{\bm k}|$ & $A_{\bm k}$ & $\Delta J_\psi$\\\hline
$(1,-1,0)$ & 2 & 1.0000 & 0.5252 & $-9.653\times10^{-2}$ & 0.4701 & 0.2351 & $-1.5612$ & 0.2195\\
$(1,0,-1)$ & 2 & 0.6180 & 0.6333 & $-1.851\times10^{-3}$ & 0.5013 & 0.2507 & $-20.243$ & 0.0629\\
$(2,-1,0)$ & 3 & 0.5000 & 0.6363 & $-1.675\times10^{-4}$ & 0.2440 & 0.1220 & $-475.05$ & 0.0091\\
$(2,0,-1)$ & 3 & 0.3090 & 0.6366 & $-7.09\times10^{-8}$ & 0.2507 & 0.1254 & $\approx-2.6\times10^{5}$ & $4\times10^{-4}$\\
\end{tabular}
\end{ruledtabular}
\caption{Primitive driving resonances ($h_{\bm k}\neq0$) with $\omega_V^*\in(0,\sqrt2)$ and $|\bm k|_1\le3$, for
$R_W=1$ ($I_W=1/2$), $F=1$, and $\varepsilon=0.01$. Widths:
$\Delta J_\psi=4\sqrt{2\varepsilon|h_{\bm k}|/|A_{\bm k}|}$, $A_{\bm k}=k_V^2\,\omega_V'(I_V^*)$.
The last three resonances lie inside the separatrix accumulation zone, where the pendulum estimate is only nominal.}
\end{table*}

%----------------------------------------------------------------------
\subsection{Parameters and primary resonances}

All quantities below are computed for the explicit model
$H_0=\tfrac12P_V^2+\tfrac12P_W^2/M_W-\tfrac12V^2+\tfrac1{12}V^4+aW-\tfrac b2W^2$,
$H_1=(W-I_{\rm eff})V$, $I_{\rm eff}=I_{\rm dc}+F\cos\nu\tau$, with the autonomous lift $\theta_3=\nu\tau$.
The frequency map $\omega_V(I_V)$, the action $I_V(E_V)$, the twist $\omega_V'(I_V)$, and the orbit harmonics
$v_{k_V}$ are evaluated both from the closed forms of Theorems~1 and 2 and by direct quadrature and orbit integration;
the two agree to better than $10^{-10}$. We use
\begin{eqnarray}
&&M_W=1,\quad a=0,\quad b=-1\ (\Omega_W=1),\quad I_{\rm dc}=0.3,\nonumber\\
&&F=1,\quad \nu=\frac{\sqrt5-1}{2}\simeq0.6180 .
\label{eq:params}
\end{eqnarray}
The drive frequency is chosen incommensurate with $\Omega_W$; the golden mean is the most badly approximable
number. The choice $b=-1$ gives a confining magnetic well. The true smallness is carried by $\varepsilon$, for which
we use values between $10^{-4}$ and $2\times10^{-2}$, as stated in each figure.

\begin{figure*}[t]
\centering
\includegraphics[width=\textwidth]{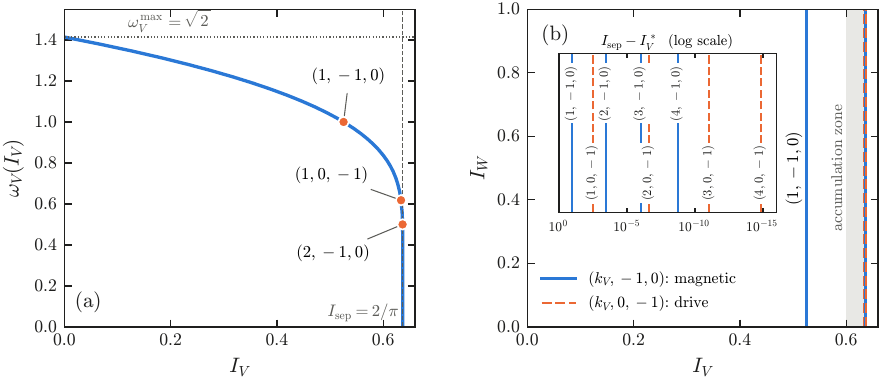}
\caption{(a) Frequency map $\omega_V(I_V)$ of the quartic well, decreasing monotonically from
$\omega_V^{\max}=\sqrt2$ at the well bottom to $0$ at the separatrix $I_{\rm sep}=2/\pi$ (Theorem~1). Dots mark the
three lowest primitive driving resonances $(1,-1,0)$, $(1,0,-1)$, and $(2,-1,0)$, at $\omega_V^*=1$, $\nu$, and $1/2$.
(b) Resonance web of the degenerate model (isochronous $W$ oscillator, $\widetilde H_0$ linear in $I_3$) in the
$(I_V,I_W)$ plane. Each resonance fixes a value $I_V=I_V^*(\bm k)$ independent of $I_W$, so the web consists of
parallel vertical lines. Solid lines: magnetic resonances $(k_V,-1,0)$; dashed lines: drive resonances $(k_V,0,-1)$.
Inset: all primitive resonances with $h_{\bm k}\neq0$ and $|\bm k|_1\le5$ on a logarithmic axis of the distance
$I_{\rm sep}-I_V^*$. All except $(1,-1,0)$ lie within $3.3\times10^{-3}$ of the separatrix (shaded accumulation zone
in the main panel), because $\omega_V\simeq2\pi/\ln(48/|E_V|)$ as $E_V\to0^-$.
Parameters as in Eq.~\eqref{eq:params}.}
\label{fig:freqmap}
\end{figure*}

The frequency map [Fig.~\ref{fig:freqmap}(a)] decreases strictly from the harmonic value $\omega_V^{\max}=\sqrt2$ at
the bottom of the well to $\omega_V\to0$ at the separatrix $I_{\rm sep}=2/\pi$, with the logarithmic flattening
$\omega_V\simeq2\pi/\ln(48/|E_V|)$ of Theorem~1. The twist $\omega_V'(I_V)<0$ therefore never vanishes in the interior.
The bounded range $\omega_V\in(0,\sqrt2)$ acts as a selection rule: only harmonics with
$|k_W\Omega_W+k_3\nu|/|k_V|<\sqrt2$ resonate. The factorized harmonics, Eqs.~(97)--(98), restrict
$k_W,k_3\in\{0,\pm1\}$ (not both nonzero), and the class $(k_V,0,0)$ is empty because it would require $\omega_V=0$.
With the golden-mean drive, the primary resonances are $(1,-1,0)$, $(1,0,-1)$, and $(2,-1,0)$ at $\omega_V^*=1$, $0.618$,
and $0.5$. All higher-order lines accumulate exponentially close to $I_{\rm sep}$ [inset of Fig.~\ref{fig:freqmap}(b)];
all of them except $(1,-1,0)$ lie within $3.3\times10^{-3}$ of the separatrix.

Table~\ref{tab:resonances} quantifies the hierarchy. The orbit harmonics decay by roughly an order of magnitude per
unit $k_V$; at $E_V=-0.4$, $|v_1|,|v_2|,|v_3|\approx0.31,\ 0.033,\ 0.0034$ (Corollary~1). The cutoff $|\bm k|_1\le5$
therefore exhausts the dynamically relevant part of the web. Two competing trends control the widths. The amplitudes
$|h_{\bm k}|$ of the three primaries are comparable ($\sim10^{-1}$), but the curvature $A_{\bm k}=k_V^2\omega_V'(I_V^*)$
[Eq.~(109)] grows rapidly toward the separatrix: from $-1.561$ for $(1,-1,0)$ deep in the well, to $-20.24$ for
$(1,0,-1)$, to $-475.0$ for $(2,-1,0)$. Consequently, at $\varepsilon=0.01$ the widths fall from
$\Delta J_\psi=0.220$ to $0.063$ and $0.009$.

The only well-separated primary is $(1,-1,0)$ at $I_V^*=0.525$. Its full width is $\Delta J_\psi=2.195\sqrt\varepsilon$,
and it sits a distance $I_{\rm sep}-I_V^*=0.111$ below the separatrix. Its upper wing reaches the separatrix when
$\tfrac12\Delta J_\psi=0.111$, i.e.\ at
\begin{equation}
\varepsilon_{\rm overlap}\simeq0.010 .
\label{eq:eps_overlap}
\end{equation}
The drive resonance $(1,0,-1)$, by contrast, lies only $3.3\times10^{-3}$ below $I_{\rm sep}$, and its layer merges
with the separatrix layer already for $\varepsilon\gtrsim10^{-4}$. For $\varepsilon<\varepsilon_{\rm overlap}$ the
primary island is isolated and the interior motion is regular. For $\varepsilon>\varepsilon_{\rm overlap}$ the island
merges with the separatrix stochastic layer, and the motion on that side becomes globally chaotic. In the degenerate
model, therefore, the onset of chaos is governed not by resonance junctions, which do not exist, but by the
layer--separatrix overlap.

\begin{figure*}[t]
\centering
\includegraphics[width=\textwidth]{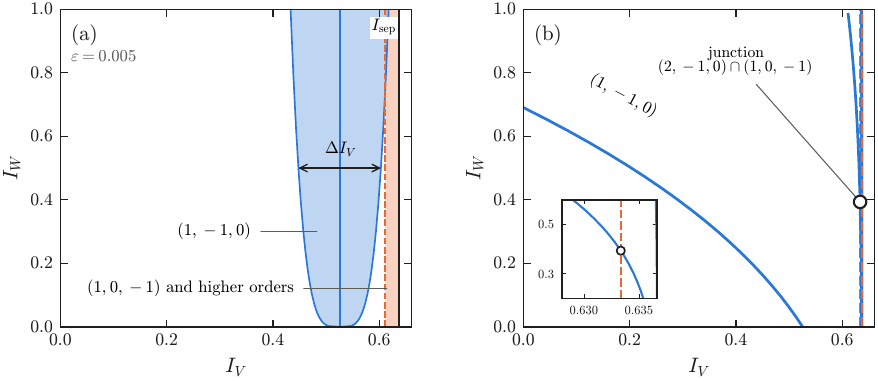}
\caption{(a) Resonance layers of the degenerate model at $\varepsilon=0.005$. Each resonance occupies the strip
$|I_V-I_V^*|\le\tfrac12\Delta I_V$, with full width
$\Delta I_V=|k_V|\Delta J_\psi=4\sqrt{2\varepsilon|h_{\bm k}(I_W)|/|\omega_V'(I_V^*)|}$ [Eq.~(91)].
For the magnetic resonance $(1,-1,0)$, $|h_{\bm k}|=\tfrac12R_W|v_1|$ with $R_W=\sqrt{2I_W}$, so the strip widens as
$I_W^{1/4}$. For the drive resonance $(1,0,-1)$, $|h_{\bm k}|=\tfrac12F|v_1|$ is independent of $I_W$; this layer and
all higher-order layers merge with the separatrix. The strips are parallel, so the degenerate web has no junctions.
At this value of $\varepsilon$ the $(1,-1,0)$ layer is isolated for $I_W\lesssim0.9$.
(b) Twist-restored model, $\Omega_W(I_W)=\Omega_W^0(1+\kappa I_W)$ with $\kappa=0.6$, $\Omega_W^0=1$.
The magnetic resonances become graphs $I_W=[k_V\omega_V(I_V)/\Omega_W^0-1]/\kappa$ (solid); the drive resonances
remain vertical (dashed). The open circle marks the transverse junction of $(2,-1,0)$ and $(1,0,-1)$ at
$(I_V^*,I_W^*)=(0.6333,\,0.3934)$ [Proposition~1(iii)]; the inset magnifies its neighbourhood.}
\label{fig:layers}
\end{figure*}

Figure~\ref{fig:layers} contrasts the degenerate and twist-restored webs. In the minimal model the layers are
parallel vertical strips [Fig.~\ref{fig:layers}(a)]. There are no double resonances, hence no organizing centers
through which a transition chain could switch from one resonance to another, and no connected Arnold web in the
strict sense. The strip widths are nevertheless not uniform: for the magnetic resonance $(1,-1,0)$ the amplitude
$|h_{\bm k}|\propto R_W=\sqrt{2I_W}$, so the layer widens as $I_W^{1/4}$. At $\varepsilon=0.005$ it is isolated for
$I_W\lesssim0.9$ and touches the separatrix zone only for larger magnetic actions.
A mild nonlinear magnetic twist, $\Omega_W(I_W)=\Omega_W^0(1+\kappa I_W)$ with $\kappa=0.6$, bends the magnetic
resonances into curves that cross the vertical drive resonances [Fig.~\ref{fig:layers}(b)]. The crossing of
$(2,-1,0)$ and $(1,0,-1)$ at $(0.6333,0.3934)$ is the junction of Proposition~1(iii). The twist makes $D_I^2H_0$ of
rank two, so resonance lines with $k_W\neq0$ acquire a transverse slope. Any theorem-level statement about
transition chains, large-gap bridging, or Nekhoroshev steepness must therefore be formulated in the twist-restored
model; the degenerate model serves as its exactly solvable skeleton.

%----------------------------------------------------------------------
\subsection{Numerical tests: resonance trapping and overlap}

\begin{figure*}[t]
\centering
\includegraphics[width=\textwidth]{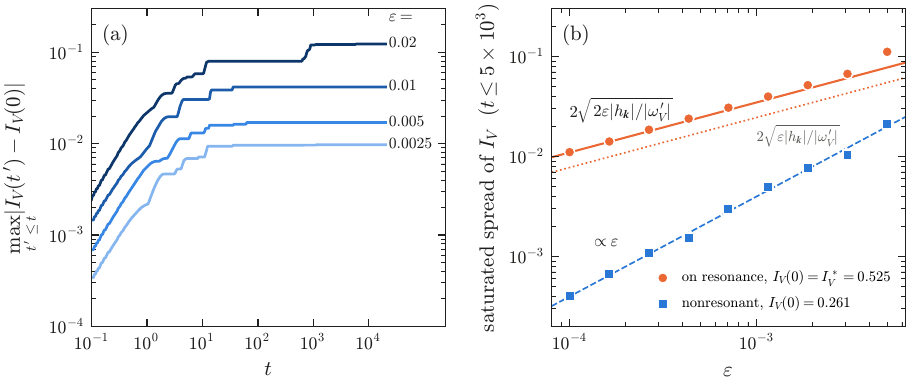}
\caption{Direct integration of the driven model, Eqs.~(35)--(36), with the parameters of Eq.~\eqref{eq:params}
(fourth-order Runge--Kutta, $\Delta\tau=0.01$). Each curve is the maximum over an ensemble of 24 initial phases
$(\theta_V,\theta_W,\theta_3)$ with $I_W(0)=1/2$.
(a) Running action spread $\max_{t'\le t}|I_V(t')-I_V(0)|$ for a nonresonant initial action $I_V(0)=0.261$
($E_V=-0.4$) and $\varepsilon=0.0025$--$0.02$. For $\varepsilon\le0.01$ the spread saturates over the integration window.
(b) Saturated spread ($t\le5\times10^3$) versus $\varepsilon$.
Squares: nonresonant initial action. The spread scales as $\varepsilon$ (dashed line, $\simeq4\varepsilon$); this is
the first-order forced excursion.
Circles: initial action on the $(1,-1,0)$ resonance, $I_V(0)=I_V^*=0.525$. The spread scales as $\sqrt\varepsilon$ and
agrees within $1\%$ at small $\varepsilon$ with the pendulum half-width
$\tfrac12\Delta I_V=2\sqrt{2\varepsilon|h_{\bm k}|/|\omega_V'|}=1.097\sqrt\varepsilon$ (solid line).
The dotted line shows the same expression without the factor $2$ from the $\pm\bm k$ pair of harmonics.
The upturn at $\varepsilon=5\times10^{-3}$ marks the onset of overlap with the $(1,0,-1)$ layer.}
\label{fig:spread}
\end{figure*}

\begin{figure*}[t]
\centering
\includegraphics[width=\textwidth]{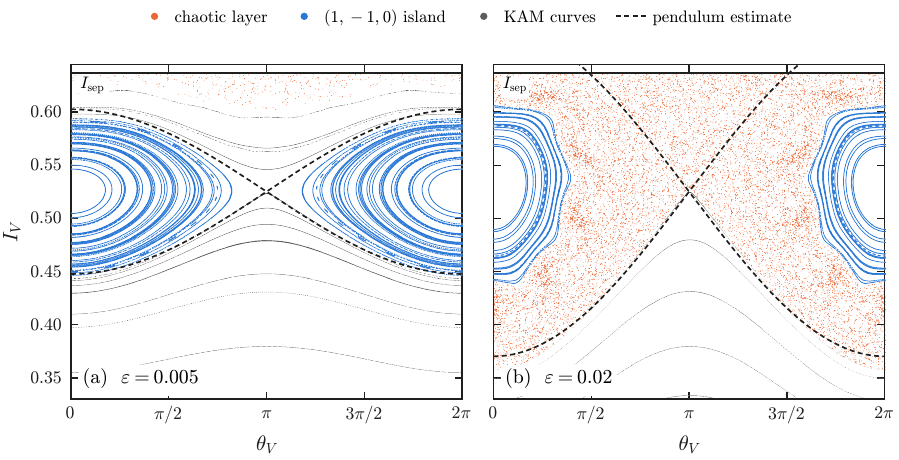}
\caption{Poincar\'e sections of the autonomous two-degree-of-freedom model ($F=0$, $I_{\rm dc}=0.3$) on the energy
shell $H=0.4$. The section is $\theta_W=0$ ($P_W=0$, $W$ maximal). Points are shown in the unperturbed
action--angle variables $(\theta_V,I_V)$ of the right well.
(a) $\varepsilon=0.005$: the primary $(1,-1,0)$ island (blue) is surrounded by rotational KAM curves (grey) that
separate it from the thin chaotic layer at the separatrix (orange). The dashed curve is the pendulum separatrix
$I_V=I_V^*\pm2\sqrt{2\varepsilon|h_{\bm k}|/|\omega_V'|}\,|\cos(\theta_V/2)|$, with no fitted parameters.
(b) $\varepsilon=0.02$: the estimated half-width ($0.155$) exceeds the gap $I_{\rm sep}-I_V^*=0.111$. The island
separatrix has dissolved into the separatrix chaotic layer, leaving only its core, in agreement with
$\varepsilon_{\rm overlap}\simeq0.010$ [Eq.~\eqref{eq:eps_overlap}].
In both panels, invariant curves below the layer bound the motion of $I_V$ (Theorem~3).}
\label{fig:poincare}
\end{figure*}

Figure~\ref{fig:spread} tests the resonant normal form by direct integration of the driven model,
Eqs.~(35)--(36). Away from resonance [$I_V(0)=0.261$], the running spread of $I_V$ saturates over the
integration window at a level $\simeq4\varepsilon$ [Fig.~\ref{fig:spread}(a) and squares in Fig.~\ref{fig:spread}(b)].
This is the first-order forced excursion: the action oscillates but does not drift. Only for $\varepsilon=0.02$,
above $\varepsilon_{\rm overlap}$, does a late step appear at $t\approx10^3$. When the trajectory starts on the
primary resonance $(1,-1,0)$, the spread is set by the island instead. It scales as $\sqrt\varepsilon$ and agrees
within $1\%$ at small $\varepsilon$ with the pendulum half-width
$\tfrac12\Delta I_V=2\sqrt{2\varepsilon|h_{\bm k}|/|\omega_V'|}=1.097\sqrt\varepsilon$ of Eq.~(91), with no fitted
parameter. This confirms both the normal form and its amplitude $2\varepsilon|h_{\bm k}|$, which comes from the
$\pm\bm k$ pair of harmonics. Without this factor the prediction, $0.776\sqrt\varepsilon$ (dotted line), would be
$30\%$ too small. The upturn at $\varepsilon=5\times10^{-3}$ marks the approach of the $(1,0,-1)$ layer: the upper
edge of the primary island ($0.603$) then nearly touches the lower edge of that layer ($0.611$).

The Poincar\'e sections of the autonomous model (Fig.~\ref{fig:poincare}) display the same island directly. At
$\varepsilon=0.005$ its separatrix coincides with the pendulum estimate, and rotational KAM curves separate it from the
thin chaotic layer at $I_{\rm sep}$, as required by Theorem~3. At $\varepsilon=0.02$, above
$\varepsilon_{\rm overlap}$ of Eq.~\eqref{eq:eps_overlap}, the island separatrix has dissolved into the separatrix
layer and only its core survives. This is the overlap transition of Eq.~\eqref{eq:overlapS} observed in the
dynamics.

These finite-time results establish resonance trapping and the overlap threshold. They neither establish nor
exclude Nekhoroshev stability or Arnold diffusion in the driven model. A numerical demonstration of Arnold diffusion
would require two things. First, a plateau of the transverse actions below the ceiling $C\varepsilon^{b}$ over times
of order $T_0e^{C\varepsilon^{-a}}$, with $a=b=1/(2n)$~[23]. Second, a slow drift along a resonance line, with
$\langle(\Delta I_\parallel)^2\rangle\simeq2D\tau$ and an exponentially small coefficient $D$. Both require
exponentially long, error-controlled integrations, which we leave for future work.

%----------------------------------------------------------------------
\subsection{Separatrix splitting: conservative versus dissipative}

\begin{figure*}[t]
\centering
\includegraphics[width=\textwidth]{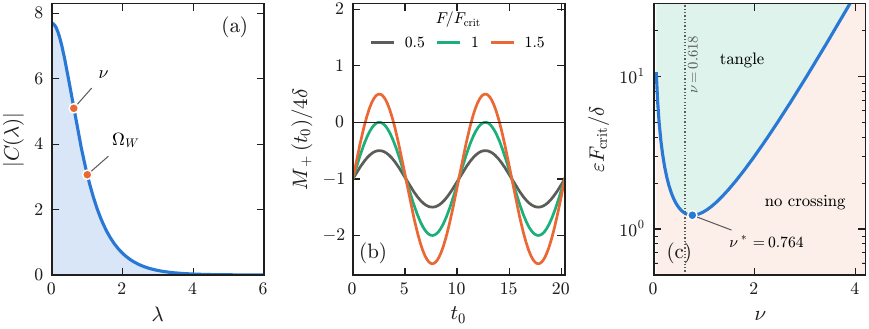}
\caption{(a) Conservative per-harmonic splitting amplitude
$|C(\lambda)|=\sqrt6\,\pi\,{\rm sech}(\pi\lambda/2)$ [Eq.~(94), Lemma~1]. It is smooth, strictly positive, and has no
threshold. Dots: the two frequencies of the model, $\nu$ and $\Omega_W$.
(b) Dissipative Melnikov function, Eq.~\eqref{eq:Mdiss}, in normalized form,
$M_+(t_0)/4\delta=-1+(F/F_{\rm crit})\sin(\nu t_0)$, for $\nu=(\sqrt5-1)/2$. Below threshold the manifolds do not
cross; at threshold they are tangent; above threshold they intersect transversally.
(c) Threshold $\varepsilon F_{\rm crit}(\nu)/\delta=4\cosh(\pi\nu/2)/(\sqrt6\,\pi\,\nu)$, Eq.~\eqref{eq:Fcrit},
separating ``tangle'' from ``no crossing''. The global minimum $1.232$ lies at $\nu^*=2x^*/\pi=0.7637$, where
$x^*\tanh x^*=1$ (Theorem~\ref{thm:dissipative}). At the golden-mean drive (dotted line) the threshold is $1.270$.
The magnetic term of Remark~\ref{rem:magnetic} lowers this threshold when $A_W>0$.}
\label{fig:melnikov}
\end{figure*}

Figure~\ref{fig:melnikov} compares the two Melnikov integrals derived in Sec.~V. The conservative per-harmonic
amplitude $C(\lambda)$ [Eq.~(94), Fig.~\ref{fig:melnikov}(a)] is a smooth, strictly positive, sech-decaying function
of the driving frequency. In the conservative lift, therefore, the stable and unstable whiskers intersect
transversally for \emph{any} nonzero forcing (Proposition~2). There is no threshold, and only the step size of the
transition-chain mechanism decreases, exponentially, at high frequency.
The dissipative, time-domain Melnikov function $M_\pm(t_0)$ [Eq.~(112), Fig.~\ref{fig:melnikov}(b)] behaves
qualitatively differently. The ferroelectric damping contributes a constant offset $-4\delta$, which shifts the
oscillatory forcing term downward. Simple zeros, and with them the homoclinic tangle, therefore exist only above the
forcing threshold
\begin{equation}
\sqrt6\,\pi\,\varepsilon F\,\nu\,{\rm sech}\Big(\frac{\pi\nu}{2}\Big)=4\delta .
\label{eq:threshold_line}
\end{equation}
Below threshold the manifolds do not cross and the dynamics relaxes. At threshold they are tangent, and above it they
intersect transversally. The critical forcing $F_{\rm crit}(\nu)$ [Fig.~\ref{fig:melnikov}(c)] is non-monotone in
$\nu$, with a unique minimum $F_{\rm crit}=1.232\,\delta/\varepsilon$ at $\nu^*=0.7637$
(Theorem~\ref{thm:dissipative}). At this optimal driving frequency the homoclinic tangle, and hence the chaotic,
excitable response of the neuromorphic oscillator, is switched on with the least electric-field amplitude. The two
descriptions are mutually consistent. The oscillatory part of $M_\pm$ is exactly $\nu|C(\nu)|\,\varepsilon F$, i.e.,
the $t_0$-derivative of the Melnikov potential, Eq.~(93). The damping enters only as the constant phase-space-area
deficit per homoclinic loop.

\section{Rigorous interpretation}
\label{sec:rigorous}

In this section we isolate those statements of the program of Sec.~III that can be proven at theorem level for the explicit Hamiltonian of Eq.~(26),
\begin{eqnarray}
&&H_0=\frac{p_V^2}{2}+\frac{p_W^2}{2}-\frac{V^2}{2}+\frac{V^4}{12}+aW-\frac{b}{2}W^2,\nonumber\\
&&H_1=\bigl(W-I_{\mathrm{eff}}\bigr)V,
\label{eq:H-explicit}
\end{eqnarray}
with $\beta=-b>0$, $\Omega_W=\sqrt{\beta}$, and, where the drive is present, $I_{\mathrm{eff}}(t)=I_{\mathrm{dc}}+F\cos\nu t$. We set $m_V=m_W=1$; all statements extend to arbitrary masses by the obvious rescaling.

%---------------------------------------------------------------------
\subsection{Twist of the quartic well}
%---------------------------------------------------------------------

\begin{theorem}[Monotone twist of the frequency map]
\label{thm:twist}
Let $E_V\in(-\tfrac34,0)$ label the orbits inside one well of $U_V(V)=-V^2/2+V^4/12$, and let $k\in(0,1)$ be the elliptic modulus defined by $E_V(k)=3(k^2-1)/(2-k^2)^2$. Then:
\begin{enumerate}
\item[(i)] the period admits the closed form
\begin{eqnarray}
T_V=2\sqrt{2-k^2}\,K(k),
\label{eq:period-closed}
\end{eqnarray}
equivalent to Eq.~(\ref{evaluated in closed form}) of the main text; \item[(ii)] $E_V(k)$ and $T_V(k)$ are strictly increasing on $(0,1)$; consequently $T_V$ is strictly increasing in $E_V$, and the frequency map $\omega_V=2\pi/T_V$ is strictly decreasing in $E_V$ and hence in the action $I_V$:
\begin{eqnarray}
&&\omega_V'(I_V)=\frac{d\omega_V}{dI_V}<0 \qquad\text{for all }\nonumber\\
&& 0<I_V<I_{\mathrm{sep}}=\frac{2}{\pi};
\end{eqnarray}
\item[(iii)] $\omega_V$ decreases from $\omega_V^{\max}=\sqrt2$ at the well bottom to $0$ at the separatrix, with the logarithmic asymptotics
\begin{eqnarray}
\omega_V \;=\; \frac{2\pi}{\ln\!\bigl(48/|E_V|\bigr)}
\;\bigl(1+o(1)\bigr),
\qquad E_V\to 0^-.
\label{eq:log-asymptotics}
\end{eqnarray}
\end{enumerate}
In particular the $V$ sector carries a nonvanishing twist on the whole
interior of the well.
\end{theorem}
\begin{proof}
(i) The in-well solution is $V_<(t)=\sqrt{6}\,(2-k^2)^{-1/2}\,\mathrm{dn}\bigl(u,k\bigr)$ with $u=(t-t_0)/\sqrt{2-k^2}$ [Eq.~(57)]. Since $\mathrm{dn}(\cdot,k)$ has period $2K(k)$, the period in $t$ is $T_V=2\sqrt{2-k^2}\,K(k)$. Writing $s=\sqrt{9+12E_V}$, one has $k^2=2s/(3+s)$, $2-k^2=6/(3+s)$, and $T_V=\sqrt{24}\,(3+s)^{-1/2}K$, which is Eq.~(40) with $r_2^2=3+s$.\\
(ii) With $x=k^2$,
\begin{widetext}
\begin{eqnarray}
\frac{dE_V}{dk}
=\frac{d}{dk}\,\frac{3(k^2-1)}{(2-k^2)^2}
=\frac{6k(2-k^2)\bigl[(2-k^2)+2(k^2-1)\bigr]}{(2-k^2)^4}
=\frac{6k^3}{(2-k^2)^3}>0 .
\end{eqnarray}
\end{widetext}
For the period, using the standard derivatives $\frac{dK}{dk}=\frac{E(k)-(1-k^2)K(k)}{k(1-k^2)}$ and abbreviating $K=K(k)$, $E=E(k)$ (complete elliptic integrals of the first and second kind),
%\begin{widetext}
\begin{eqnarray}
&&\frac{dT_V}{dk}=\frac{-2k}{\sqrt{2-k^2}}\,K+\nonumber\\
&&2\sqrt{2-k^2}\frac{E-(1-k^2)K}{k(1-k^2)}\\
&&=\frac{2}{k(1-k^2)\sqrt{2-k^2}}\;g(k),\nonumber
\end{eqnarray}
%\end{widetext}
with
\begin{equation}
g(k)\equiv(2-k^2)\,E(k)-2(1-k^2)\,K(k).
\label{eq:gk}
\end{equation}
It remains to show $g>0$ on $(0,1)$. Clearly $g(0)=2\cdot\tfrac\pi2-2\cdot\tfrac\pi2=0$. Differentiating and using $\frac{dE}{dk}=\frac{E-K}{k}$,
%\begin{widetext}
\begin{eqnarray}
&&k\,g'(k)
=-2k^2E+(2-k^2)(E-K)+\nonumber\\
&&4k^2K-2E+2(1-k^2)K=\\
&&3k^2\,\bigl(K(k)-E(k)\bigr)>0,\nonumber
\end{eqnarray}
%\end{widetext}
for $k\in(0,1)$, since $K>E$ there. Hence $g(k)>0$ for $k>0$, so $T_V$ is strictly increasing in $k$ and therefore in $E_V$. Finally $dI_V/dE_V=1/\omega_V>0$, so 
\begin{equation*}
\omega_V'(I_V)=\omega_V\frac{d\omega_V}{dE_V}=-\frac{4\pi^2}{T_V^3}\frac{dT_V}{dE_V}<0,
\end{equation*}
which is Eq.~(108) with a definite sign.\\
(iii) At the bottom, $k\to0$ gives $T_V\to\pi\sqrt2$, i.e. $\omega_V\to\sqrt2=\sqrt{U_V''(\pm\sqrt3)}$. Near the separatrix put $k'^2=1-k^2\to0$; then $E_V=-3k'^2(1+O(k'^2))$, i.e. $k'=\sqrt{|E_V|/3}\,(1+O(|E_V|))$, while $K(k)=\ln(4/k')+O(k'^2\ln k')$ and $\sqrt{2-k^2}\to1$. Hence $T_V=2\ln(4/k')+o(1)=\ln(48/|E_V|)+o(\ln|E_V|)$, which gives Eq.~(\ref{eq:log-asymptotics}).
\end{proof}
\begin{remark}
Theorem~\ref{thm:twist} is the analytic backbone of Fig.~1(a): the numerically tabulated frequency map is monotone \emph{by theorem}, not by inspection, and the accumulation of resonance lines at $I_{\mathrm{sep}}$ in Fig.~1(b) is quantified by Eq.~(\ref{eq:log-asymptotics}): the resonant action $I_V^*(\bm k)$ approaches $I_{\mathrm{sep}}$ exponentially fast in $1/\omega_V^*(\bm k)$.
\end{remark}

%---------------------------------------------------------------------
\subsection{Exact harmonic content and selection rules}
%---------------------------------------------------------------------

\begin{theorem}[Harmonic spectrum of the perturbation]
\label{thm:harmonics}
Let $q=e^{-\pi K'(k)/K(k)}$ be the elliptic nome of the in-well orbit at energy $E_V(k)$, where $K'(k)=K(\sqrt{1-k^2})$. Then the Fourier coefficients of Eq.~(55) are, exactly,
\begin{eqnarray}
&&v_0=\frac{\sqrt6\,\pi}{2\sqrt{2-k^2}\,K(k)},\\
&&v_n=v_{-n}=\frac{\sqrt6\,\pi}{\sqrt{2-k^2}\,K(k)}\;\frac{q^{\,n}}{1+q^{\,2n}},\quad n\ge1 .\nonumber
\label{eq:vn-exact}
\end{eqnarray}
Consequently:
\begin{enumerate}
\item[(i)] $|v_n|$ decays exponentially, $|v_n|\le C(k)\,q^{\,n}$, at the rate set by the nome, and the ratio of consecutive harmonics satisfies $v_{n+1}/v_n\to q$;
\item[(ii)] the full harmonics factorize as in Eq.~(53)--(54), and
$h_{\bm k}=0$ unless
\begin{eqnarray}
&&k_W\in\{0,\pm1\},\qquad k_3\in\{0,\pm1\},\qquad \text{not both } \nonumber\\
&&|k_W|=|k_3|=1;
\end{eqnarray}
\item[(iii)] no driving resonance of class $(k_V,0,0)$ exists in the interior of the well: for $k_V\neq0$ the condition
$k_V\,\omega_V(I_V)=0$ has no solution with
$0<I_V<I_{\mathrm{sep}}$.
\end{enumerate}
\end{theorem}

\begin{proof}
The orbit is $V(\theta_V)=\sqrt6\,(2-k^2)^{-1/2}\,\mathrm{dn}(u,k)$ with $u=t/\sqrt{2-k^2}$; since $T_V=2\sqrt{2-k^2}\,K$ (Theorem~\ref{thm:twist}), the angle is $\theta_V=\omega_V t=\pi u/K$. The classical Fourier expansion of $\mathrm{dn}$ (see, e.g., Whittaker--Watson or Abramowitz--Stegun
16.23.3) reads
\begin{eqnarray}
&&\mathrm{dn}(u,k)=\frac{\pi}{2K}
+\frac{2\pi}{K}\sum_{n\ge1}\frac{q^{\,n}}{1+q^{\,2n}}
\cos\!\Bigl(\frac{n\pi u}{K}\Bigr)=\nonumber\\
&&\frac{\pi}{2K}+\frac{2\pi}{K}\sum_{n\ge1}\frac{q^{\,n}}{1+q^{\,2n}}\cos n\theta_V .
\end{eqnarray}
Multiplying by $\sqrt6\,(2-k^2)^{-1/2}$ and passing to exponential coefficients, $v_{\pm n}=\tfrac12\times$(cosine coefficient), gives Eq.~(\ref{eq:vn-exact}). Part (i) is immediate since
$1/2\le(1+q^{2n})^{-1}\le1$.

(ii) In the perturbation $H_1=V(\theta_V)\bigl[W(\theta_W)-I_{\mathrm{dc}}-F\cos\theta_3\bigr]$ the factors $W(\theta_W)=W_c+R_W\cos\theta_W$ and $F\cos\theta_3$ are pure first harmonics; by orthogonality of $\{e^{ik\theta}\}$ on $\mathbb T^1$ their Fourier support is $k_W\in\{0,\pm1\}$ and $k_3\in\{0,\pm1\}$ respectively, and since $H_1$ is a \emph{sum} of a $(\cdot,k_W,0)$ family and a $(\cdot,0,k_3)$ family, mixed terms with $|k_W|=|k_3|=1$ carry zero coefficient. The triple integral of Eq.~(42) then factorizes into Eq.~(53)--(54).

(iii) By Theorem~\ref{thm:twist}, $\omega_V(I_V)\in(0,\sqrt2)$ is strictly positive on the open interval, so $k_V\,\omega_V=0$ forces $k_V=0$.
\end{proof}

\begin{corollary}[Quantitative check]
\label{cor:nome}
At $E_V=-0.4$ one finds $k^2=0.8117$, $K=2.295$, $K'=1.655$, hence $q=e^{-\pi K'/K}=0.103$, and Eq.~(\ref{eq:vn-exact}) gives
\begin{eqnarray}
v_1=0.315,\qquad v_2=0.033,\qquad v_3=0.0034,
\end{eqnarray}
reproducing the numerically quoted values $|v_1|,|v_2|,|v_3|\approx0.31,\,0.033,\,0.0034$ and identifying the empirical ``order of magnitude per unit $k_V$'' decay as the  $q(E_V)$. The cutoff $|\bm k|_1\le5$ used in Figs.~1--2 therefore discards only harmonics suppressed at least as $q^4\sim10^{-4}$.
\end{corollary}

%---------------------------------------------------------------------
\subsection{Confinement in the autonomous model}
%---------------------------------------------------------------------

\begin{theorem}[Kolmogorov degeneracy, iso-energetic non-degeneracy, and perpetual confinement]
\label{thm:confinement}
Consider the autonomous two-degree-of-freedom system $H=H_0+\varepsilon H_1$ of Eq.~(\ref{eq:H-explicit}) with constant $I_{\mathrm{eff}}=I_{\mathrm{dc}}$ and $\beta>0$, written in action--angle variables,
\begin{eqnarray}
H_0(I_V,I_W)=E_V(I_V)+\Omega_W I_W-\frac{a^2}{2\beta}.
\end{eqnarray}
Then, on every compact subinterval $I_V\in[\delta,I_{\mathrm{sep}}-\delta]$, $\delta>0$:
\begin{enumerate}
\item[(i)] the Kolmogorov determinant vanishes identically,
\begin{eqnarray}
\det D^2_I H_0
=\det\begin{pmatrix}\omega_V'(I_V)&0\\[2pt]0&0\end{pmatrix}=0 ,
\end{eqnarray}
so the Kolmogorov non-degeneracy condition Eq.~(7) fails everywhere; \item[(ii)] nevertheless the iso-energetic (Arnold) determinant Eq.~(8) is nonzero everywhere:
\begin{equation}
\det\begin{pmatrix}
\omega_V' & 0 & \omega_V\\
0 & 0 & \Omega_W\\
\omega_V & \Omega_W & 0
\end{pmatrix}
=-\,\Omega_W^2\,\omega_V'(I_V)\;>\;0 ,
\label{eq:isoenergetic}
\end{equation}
by Theorem~\ref{thm:twist};
\item[(iii)] consequently, for every energy $h$ in the corresponding range there exists $\varepsilon_0(\delta)>0$ such that for $0<\varepsilon<\varepsilon_0$ a set of positive measure of the
unperturbed two-tori with Diophantine frequency ratio $\omega_V/\Omega_W$ persists on the energy surface
$\{H=h\}$ (iso-energetic KAM theorem \cite{Arnold63,broer1996quasi}), and every persisting torus is of codimension one in the three-dimensional energy surface; \item[(iv)] therefore the actions of \emph{every} orbit on $\{H=h\}$ starting between two adjacent persisting tori remain confined for all times,
\begin{eqnarray}
\sup_{t\in\mathbb R}\bigl|I(t)-I(0)\bigr|\le C\sqrt{\varepsilon},
\end{eqnarray}
i.e. Arnold diffusion is impossible in the autonomous two-degree-of-freedom FM-FE model, for reasons of topology rather than of smallness of $\varepsilon$.
\end{enumerate}
\end{theorem}

\begin{proof}
(i) $\Omega_W$ is independent of both actions (isochrony of the harmonic $W$ well), so the second row and column of the Hessian vanish.

(ii) Expanding the bordered determinant along its second row, the only nonzero entry is $\Omega_W$ in position $(2,3)$, with cofactor $(-1)^{2+3}\det\begin{psmallmatrix}\omega_V'&0\\ \omega_V&\Omega_W
\end{psmallmatrix}=-\omega_V'\Omega_W$; hence the determinant equals $-\Omega_W^2\omega_V'>0$ since $\omega_V'<0$ (Theorem~\ref{thm:twist}) and $\Omega_W=\sqrt\beta>0$.

(iii) On the compact set $[\delta,I_{\mathrm{sep}}-\delta]$ the Hamiltonian $H_0$ is real-analytic in $I$ and the perturbation $H_1$ is real-analytic in $(I,\theta)$ (the in-well elliptic solutions are analytic and the set avoids the separatrix). Iso-energetic non-degeneracy (ii) is precisely the hypothesis of the iso-energetic KAM theorem: on each energy surface, all unperturbed tori whose frequency ratio $[\omega_V:\Omega_W]$ is Diophantine, of relative measure $1-O(\sqrt\varepsilon)$, survive the perturbation as invariant Lagrangian two-tori, slightly deformed. A two-torus inside a three-dimensional energy surface has codimension one.

(iv) A codimension-one invariant torus separates the energy surface locally into two components. Order the surviving tori by their $I_V$ value; the gaps between adjacent Diophantine tori have width
$O(\sqrt\varepsilon)$ by the measure estimate in (iii). An orbit starting in a gap can never cross the bounding tori (invariance), so its action excursion is bounded by the gap width for all $t\in\mathbb R$, positive and negative. This is the standard two-degree-of-freedom trapping argument \cite{LichtenbergLieberman1992}; no assumption on the structure of the resonances inside the gap is
needed.
\end{proof}

\begin{remark}
Theorem~\ref{thm:confinement} sharpens Remark~1 of the Introduction into a rigorous statement, and it does so despite the failure of the Kolmogorov condition: it is the \emph{iso-energetic} determinant
Eq.~(\ref{eq:isoenergetic}), saved by the twist of the quartic well, that carries the burden. It also explains why, in Eq.~(49), the vanishing of $\det D^2_I H_0$ does not contradict the confinement observed numerically in Fig.~3(a): confinement in $2$ DOF requires iso-energetic non-degeneracy only. In the driven (three-frequency) lift the tori become codimension two and the argument of item (iv) collapses---this is exactly the door through which Arnold diffusion can enter.
\end{remark}

%---------------------------------------------------------------------
\subsection{Geometry of the resonance web}
%---------------------------------------------------------------------

\begin{proposition}[Web geometry: degenerate and twist-restored]
\label{prop:web}
Consider the lifted model with frequency vector $\bigl(\omega_V(I_V),\,\Omega_W,\,\nu\bigr)$ and driving resonances $f_{\bm k}(I)\equiv k_V\omega_V(I_V)+k_W\Omega_W(I_W)+k_3\nu=0$, restricted to the classes of Theorem~\ref{thm:harmonics}(ii) with
$k_V\neq0$.
\begin{enumerate}
\item[(i)] (\emph{Degenerate model}, $\Omega_W=\mathrm{const}$.) By strict monotonicity of $\omega_V$ each resonance is a single vertical line $I_V=I_V^{*}(\bm k)$ in the $(I_V,I_W)$ plane, and it exists iff $|k_W\Omega_W+k_3\nu|/|k_V|<\sqrt2$. Two distinct primitive driving resonances never intersect: the web has no junctions.
\item[(ii)] (\emph{Twist-restored model}, $\Omega_W(I_W)=\Omega_W^0(1+\kappa I_W)$, $\kappa\neq0$.)
Resonances with $k_W\neq0$ become graphs of strictly monotone $C^1$ curves $I_W=I_W^{\bm k}(I_V)$. Two resonances $\bm k$, $\bm k'$ intersect transversally wherever they meet:
\begin{eqnarray}
&&\det\begin{pmatrix}
k_V\,\omega_V' & k_W\,\kappa\,\Omega_W^0\\
k_V'\,\omega_V' & k_W'\,\kappa\,\Omega_W^0
\end{pmatrix}
=\\
&&\kappa\,\Omega_W^0\,\omega_V'\,\bigl(k_Vk_W'-k_V'k_W\bigr) \neq0\nonumber
\end{eqnarray}
iff the projections $(k_V,k_W)$ and $(k_V',k_W')$ are not parallel. \item[(iii)] (\emph{Explicit junction.}) For the parameters of the figures ($\Omega_W^0=1$, $\nu=(\sqrt5-1)/2$, $\kappa=0.6$) the
resonances $(2,-1,0)$ and $(1,0,-1)$ intersect transversally at the unique point
\begin{eqnarray}
&&I_V^{*}=\omega_V^{-1}(\nu)=0.6333,\\
&&I_W^{*}=\frac{2\nu-\Omega_W^0}{\kappa\,\Omega_W^0}=0.3934 ,\nonumber
\end{eqnarray}
which is the junction visible in Fig.~2(b). More generally such a junction exists iff $0<(|k_V|\nu/|k_W|-\Omega_W^0)/(\kappa\Omega_W^0)$ lies in the admissible $I_W$ range.
\end{enumerate}
\end{proposition}

\begin{proof} (i) For fixed $(k_W,k_3)$ the resonance condition reads $\omega_V(I_V)=-(k_W\Omega_W+k_3\nu)/k_V$; by Theorem~\ref{thm:twist} the map $\omega_V$ is a strictly decreasing bijection of $(0,I_{\mathrm{sep}})$ onto $(0,\sqrt2)$, so a solution $I_V^*$ exists iff the right-hand side lies in $(0,\sqrt2)$, and it is unique; the condition contains no $I_W$, so the locus is a vertical line. Two vertical lines either coincide or are disjoint; they coincide only if the resonance values agree, which for primitive $\bm k\neq\bm k'$ and $\nu\notin\mathbb Q\,\Omega_W$ (guaranteed by the golden-mean choice) forces $\bm k=\bm k'$.

(ii) Solve $f_{\bm k}=0$ for $I_W$: $I_W^{\bm k}(I_V)=-\bigl(k_V\omega_V(I_V)+k_3\nu+k_W\Omega_W^0\bigr)
/(k_W\kappa\Omega_W^0)$, a $C^1$ function with $dI_W/dI_V=-k_V\omega_V'/(k_W\kappa\Omega_W^0)\neq0$. The gradient of $f_{\bm k}$ is $\bigl(k_V\omega_V',\,k_W\kappa\Omega_W^0\bigr)$, and the displayed determinant of the two gradients is nonzero exactly when $k_Vk_W'\neq k_V'k_W$; nonvanishing of the determinant of gradients at a common zero is the definition of transversal intersection of the two level curves.

(iii) The resonance $(1,0,-1)$ is the vertical line $\omega_V(I_V)=\nu$, i.e. $I_V^*=0.6333$ from Table~I. Substituting into $(2,-1,0)$: $2\nu=\Omega_W^0(1+\kappa I_W)$ gives the stated $I_W^*$; with $\nu=0.6180$, $\kappa=0.6$ this is $I_W^*=0.2361/0.6=0.3934\in(0,1)$. Transversality follows from (ii)
with $(k_V,k_W)=(2,-1)$, $(k_V',k_W')=(1,0)$: $k_Vk_W'-k_V'k_W=0-(-1)=1\neq0$.
\end{proof}

%---------------------------------------------------------------------
\subsection{Separatrix splitting: conservative and dissipative}
%---------------------------------------------------------------------

\begin{lemma}[Melnikov integrals of the $V$ saddle]
\label{lem:integrals}
For the homoclinic orbit $V_s(t)=s\sqrt6\,\mathrm{sech}\,t$, $s=\pm1$, and any $\lambda>0$,
\begin{eqnarray}
&&\int_{-\infty}^{\infty} V_s(t)\cos\lambda t\;dt
= s\,\frac{\sqrt6\,\pi}{\cosh(\pi\lambda/2)} = C(\lambda),\nonumber\\
&&\int_{-\infty}^{\infty}\dot V_s^2(t)\,dt = 4 .
\label{eq:melnikov-integrals}
\end{eqnarray}
\end{lemma}

\begin{proof} For the first integral it suffices to evaluate $\int_{-\infty}^\infty \mathrm{sech}\,t\;e^{i\lambda t}dt$. The integrand is meromorphic with simple poles of $\mathrm{sech}$ at
$t=i\pi(n+\tfrac12)$, $n\in\mathbb Z$. Closing the contour with the rectangle of height $\pi$ (vertices $\pm R$, $\pm R+i\pi$) and using $\mathrm{sech}(t+i\pi)=-\mathrm{sech}\,t$, the integral over the top
edge equals $e^{-\pi\lambda}$ times the original one, while the residue at $t=i\pi/2$, where $\mathrm{Res}\,\mathrm{sech}=-i$, contributes $2\pi i\cdot(-i)e^{-\pi\lambda/2}=2\pi
e^{-\pi\lambda/2}$. Hence $(1+e^{-\pi\lambda})\int=2\pi e^{-\pi\lambda/2}$, i.e. $\int=\pi\,\mathrm{sech}(\pi\lambda/2)$, and the first formula follows after multiplying by $s\sqrt6$ (the sine part vanishes by parity). For the second, $\dot V_s=-s\sqrt6\,\mathrm{sech}\,t\tanh t$, so $\int\dot V_s^2\,dt=6\int\mathrm{sech}^2t\tanh^2t\,dt =6\bigl[\tfrac13\tanh^3t\bigr]_{-\infty}^{\infty}=4$.
\end{proof}

\begin{proposition}[No threshold in the conservative lift]
\label{prop:conservative}
In the conservative three-frequency model $\widetilde H_0+\varepsilon H_1$, the Melnikov potential of Eq.~(47),
\begin{eqnarray}
L(\phi_W,\phi_3)=-A_W\,C(\Omega_W)\cos\phi_W+F\,C(\nu)\cos\phi_3 ,
\end{eqnarray}
has, for any $A_W>0$, $F>0$, exactly four critical points on $\mathbb T^2$, all nondegenerate, located at
$(\phi_W,\phi_3)\in\{0,\pi\}^2$. Consequently, for all sufficiently small $\varepsilon>0$ the whiskers of the invariant tori intersect transversally, $W^u\pitchfork W^s$, and the splitting has no forcing threshold: it is nonzero for arbitrarily weak drive, decaying only as $\mathrm{sech}(\pi\lambda/2)$ with the driving frequency
(Lemma~\ref{lem:integrals}).
\end{proposition}

\begin{proof} $C(\lambda)\neq0$ for every finite $\lambda$ by Lemma~\ref{lem:integrals}, so both amplitudes
$a\equiv A_WC(\Omega_W)$ and $b\equiv FC(\nu)$ are nonzero. Critical points require $\sin\phi_W=\sin\phi_3=0$, giving the four points stated: 
The Hessian there is ${\rm diag}(a\cos\phi_W,\,-b\cos\phi_3)$, with determinant $-ab\cos\phi_W\cos\phi_3\neq0$: saddles at $(0,0)$ and $(\pi,\pi)$, a minimum at $(0,\pi)$, a maximum at $(\pi,0)$.''
Nondegenerate critical points of the Melnikov potential correspond, at first order in $\varepsilon$ and by the implicit function theorem, to transverse homoclinic/heteroclinic intersections of the perturbed whiskers
\cite{wiggins2003introduction}.
\end{proof}

\begin{theorem}[Dissipative threshold and optimal driving frequency]\label{thm:dissipative}
Consider Eq.~(35) with the phenomenological damping $-\delta\,dV/d\tau$, $\delta=O(\varepsilon)>0$,
and with the magnetic mode at rest, $W\equiv W_c$ ($A_W=0$).
Along the homoclinic orbits $V_s(t)=s\sqrt6\,\operatorname{sech}t$, $s=\pm1$, the Melnikov function is
\begin{equation}
M_\pm(t_0)=-4\delta\pm\sqrt6\,\pi\,\varepsilon F\,\nu\,
\operatorname{sech}\!\Big(\frac{\pi\nu}{2}\Big)\sin(\nu t_0).
\label{eq:Mdiss}
\end{equation}
Define
\begin{equation}
F_{\rm crit}(\nu)=\frac{4\delta}{\sqrt6\,\pi\,\varepsilon}\,\frac{\cosh(\pi\nu/2)}{\nu}.
\label{eq:Fcrit}
\end{equation}
Then:
\begin{enumerate}
\item[(i)] If $F>F_{\rm crit}(\nu)$, $M_\pm$ has simple zeros. For $\varepsilon$ sufficiently small the
stable and unstable manifolds of the perturbed saddle then intersect transversally. This produces a homoclinic tangle
and, by the Smale--Birkhoff theorem, chaotic (horseshoe) dynamics.
If $F<F_{\rm crit}(\nu)$, $M_\pm$ has no zeros and the manifolds do not intersect.
At $F=F_{\rm crit}(\nu)$ they are tangent at first order.
\item[(ii)] $F_{\rm crit}$ attains a unique global minimum on $(0,\infty)$ at
\begin{equation}
\nu^*=\frac{2x^*}{\pi}\simeq0.7637,\qquad x^*\tanh x^*=1,\quad x^*\simeq1.19968,
\label{eq:nustar}
\end{equation}
with minimal value
\begin{equation}
F_{\rm crit}(\nu^*)=\frac{4\delta\cosh x^*}{\sqrt6\,\pi\,\varepsilon\,\nu^*}\simeq1.232\,\frac{\delta}{\varepsilon}.
\label{eq:Fmin}
\end{equation}
The threshold diverges as $4\delta/(\sqrt6\pi\varepsilon\nu)$ for $\nu\to0^+$ and as
$2\delta\,e^{\pi\nu/2}/(\sqrt6\pi\varepsilon\nu)$ for $\nu\to\infty$.
\end{enumerate}
\end{theorem}

\begin{proof}
(i) With $\tau\to t$ and the drive phase shifted by $t_0$, the perturbed equation reads
$\ddot V=V-V^3/3+\varepsilon\,[I_{\rm dc}-W_c+F\cos\nu(t+t_0)]-\delta\dot V$.
The Melnikov function is the work done by the perturbation along the unperturbed homoclinic orbit,
\begin{align*}
M(t_0)&=\int_{-\infty}^{\infty}\dot V_s(t)\Big\{\varepsilon\big[I_{\rm dc}-W_c+F\cos\nu(t+t_0)\big] \\
&-\delta\,\dot V_s(t)\Big\}\,dt .
\end{align*}
%\begin{equation*}
%M(t_0)=\int_{-\infty}^{\infty}\dot V_s(t)\Big\{\varepsilon\big[I_{\rm dc}-W_c+F\cos\nu(t+t_0)\big]-\delta\,\dot V_s(t)\Big\}\,dt .
%\end{equation*}
The constant force does no work, since $\int\dot V_s\,dt=V_s(+\infty)-V_s(-\infty)=0$.
By Lemma~1, Eq.~(148), the damping contributes $-\delta\int\dot V_s^2\,dt=-4\delta$.
For the drive term we write $\cos\nu(t+t_0)=\cos\nu t\cos\nu t_0-\sin\nu t\sin\nu t_0$.
Since $\dot V_s$ is odd, the $\cos\nu t$ part vanishes. Integrating by parts in the $\sin\nu t$ part gives
$\int\dot V_s\sin\nu t\,dt=-\nu\int V_s\cos\nu t\,dt=-\nu\,C(\nu)$, where
$C(\nu)=s\sqrt6\,\pi\operatorname{sech}(\pi\nu/2)$ by Lemma~1. Hence
\begin{equation*}
M(t_0)=-4\delta+\varepsilon F\,\nu\,C(\nu)\sin\nu t_0 ,
\end{equation*}
which is Eq.~\eqref{eq:Mdiss} with the sign $\pm$ given by $s$.
Equivalently, the oscillatory part is $\varepsilon F\,\partial_{t_0}$ of the drive part of the Melnikov potential, Eq.~(93).

Put $A(\nu,F)=\sqrt6\,\pi\,\varepsilon F\,\nu\operatorname{sech}(\pi\nu/2)\ge0$, so that
$M_\pm(t_0)=-4\delta\pm A\sin\nu t_0$ is periodic with range $[-4\delta-A,\,-4\delta+A]$.
Zeros exist iff $A\ge4\delta$, i.e.\ $F\ge F_{\rm crit}(\nu)$.
For $A>4\delta$ the zeros satisfy $\sin\nu t_0=\pm4\delta/A$, hence
$|\cos\nu t_0|=\sqrt{1-(4\delta/A)^2}\neq0$ and $M_\pm'(t_0)=\pm A\nu\cos\nu t_0\neq0$: the zeros are simple.
For $A<4\delta$, $M_\pm<0$ for all $t_0$. For $A=4\delta$ the zeros are double ($M_\pm=M_\pm'=0$).
The standard Melnikov theorem for dissipative periodic perturbations~[24] then gives the three alternatives:
a transverse intersection for simple zeros, no intersection for sign-definite $M$, and a quadratic tangency at threshold.

(ii) Minimizing Eq.~\eqref{eq:Fcrit} amounts to minimizing $h(\nu)=\cosh(\pi\nu/2)/\nu$.
With $x=\pi\nu/2$ we have $h=\tfrac{\pi}{2}\cosh x/x$ and
\begin{equation*}
\frac{d}{dx}\frac{\cosh x}{x}=\frac{x\sinh x-\cosh x}{x^2}=\frac{\cosh x}{x^2}\,\big(x\tanh x-1\big).
\end{equation*}
The function $\varphi(x)=x\tanh x$ is strictly increasing on $(0,\infty)$, being a product of two positive strictly increasing
functions, with $\varphi(0^+)=0$ and $\varphi(\infty)=\infty$. Therefore $\varphi(x)=1$ has a unique root;
numerically $x^*=1.19968$, so $\nu^*=2x^*/\pi=0.7637$.
Since $x\tanh x-1<0$ for $x<x^*$ and $>0$ for $x>x^*$, $h$ decreases and then increases, and $\nu^*$ is the unique global minimum.
Evaluating gives $F_{\rm crit}(\nu^*)=4\delta\cosh(x^*)/(\sqrt6\pi\varepsilon\nu^*)=1.232\,\delta/\varepsilon$.
The limits follow from $\cosh x\to1$ ($x\to0$) and $\cosh x\sim e^x/2$ ($x\to\infty$).
\end{proof}

\begin{remark}[Magnetic contribution]\label{rem:magnetic}
For $A_W>0$ the magnetic mode enters the force on $V$ through $-\varepsilon W$. To leading order it may be taken along its
unperturbed motion $W(t)=W_c+A_W\cos(\Omega_Wt+\phi_W)$: the back-action $-\varepsilon V$ in the $W$ equation changes $W$
by $O(\varepsilon)$, i.e.\ the force on $V$ by $O(\varepsilon^2)$. The same integration by parts as in the proof then gives
\begin{eqnarray}
&&M(t_0)=-4\delta+\varepsilon F\nu\,C(\nu)\sin\nu t_0-\\
&&\varepsilon A_W\Omega_W\,C(\Omega_W)\sin(\Omega_Wt_0+\phi_W).\nonumber
\label{eq:Mfull}
\end{eqnarray}
For $\nu/\Omega_W\notin\mathbb Q$ this function is quasi-periodic in $t_0$. By Kronecker's theorem the two phases fill the torus densely, so
\begin{eqnarray}
&&\sup_{t_0}M=-4\delta+\varepsilon\big[F\nu|C(\nu)|+A_W\Omega_W|C(\Omega_W)|\big],\nonumber\\
&&\inf_{t_0}M=-\varepsilon\big[F\nu|C(\nu)|+A_W\Omega_W|C(\Omega_W)|\big]\\
&&-4\delta<0 \nonumber.
\end{eqnarray}
Hence $M$ changes sign, and has zeros, iff $F>F_{\rm crit}^{(A_W)}$, where
\begin{equation}
F_{\rm crit}^{(A_W)}(\nu)=\frac{4\delta/\varepsilon-A_W\Omega_W|C(\Omega_W)|}{\nu\,|C(\nu)|}.
\label{eq:FcritAW}
\end{equation}
Zeros with $M'(t_0)\neq0$ yield transverse intersections, by the Melnikov theorem for quasi-periodic perturbations~[24].
For $F<F_{\rm crit}^{(A_W)}$, $M<0$ and the manifolds do not intersect.
The magnetic precession therefore lowers the electric threshold. If
$A_W\Omega_W|C(\Omega_W)|\ge4\delta/\varepsilon$, the precession alone produces the tangle, without electric driving.
\end{remark}

\begin{remark}\label{rem:optimal}
Theorem~\ref{thm:dissipative}(ii) is the analytic content of Fig.~6(c). The boundary between ``tangle'' and ``no crossing''
in the $(\nu,F)$ plane has a unique minimum at $\nu^*\approx0.764$. At the golden-mean drive $\nu=(\sqrt5-1)/2\approx0.618$ used in
the numerics the threshold is $1.270\,\delta/\varepsilon$, only $3\%$ above the minimum.
Physically, $\nu^*$ is the driving frequency at which the homoclinic tangle, i.e.\ the chaotic excitable response of the
FM--FE neuromorphic oscillator, is switched on with the least electric-field amplitude at fixed ferroelectric damping.
Together, Proposition~2 and Theorem~\ref{thm:dissipative} make rigorous the contrast displayed in Fig.~6(a,b).
The conservative splitting has no threshold, whereas the dissipative splitting does.
Both share the same oscillatory content: the forcing part of the Melnikov function is $\partial_{t_0}$ of the Melnikov potential.
\end{remark}

\section{Conclusions}
\label{sec:conclusions}

We have developed a conservative oscillator description of a magnetoelectrically coupled ferromagnetic--ferroelectric system. The local magnetic reduction expresses one transverse spin component as a coordinate and the other as its conjugate momentum, so that the effective kinetic term follows from ordinary magnetic precession. Its physical application requires a stable magnetic reference state and transverse amplitudes within the local expansion. The polarization can supply a voltage or current to an external artificial neuron, with the measured signal determined by screening,
capacitance, leakage, and electrical loading.

The principal analytical results concern the integrable frequency map and its perturbation. The quartic well has a strictly decreasing in-well frequency, nonzero twist, and explicit elliptic-function Fourier coefficients. Together with the harmonic magnetic mode, these quantities determine the leading supported resonance families and their local pendulum widths. In the autonomous model the isoenergetic determinant is nonzero even though the Kolmogorov determinant vanishes, permitting KAM confinement between surviving tori on regular compact energy domains.

Periodic electric forcing removes the codimension-one torus barrier of the autonomous problem. The minimal lifted frequency map still gives parallel resonance loci, whereas an explicitly defined nonlinear magnetic extension creates transverse resonance intersections. These geometric observations neither prove nor exclude Arnold diffusion. In particular, a junction does not by itself provide a transition chain, and the pre-existing polarization
separatrix requires a separate analysis of its invariant manifolds and scattering dynamics.

The homoclinic Fourier integral yields local conservative splitting and, after adding weak damping with a consistent normalization, a leading-order threshold for a specified scalar reduction. The full coupled problem includes an additional magnetic term in the Melnikov function. The corrected resonance-width estimate provides only a geometric comparison with the unperturbed separatrix; it does not establish a sharp global-chaos threshold.
Likewise, the reported finite-time action excursions do not verify a Nekhoroshev theorem or an exponential diffusion law.

The outstanding transport question is whether the coupled system admits substantial action drift through sequences of homoclinic excursions or resonant transitions. Resolving it requires compatible scattering and internal dynamics, or reproducible long-time simulations with controlled errors. Establishing neuromorphic operation additionally requires a calibrated physical parameter set and a loaded, dissipative readout model. The present results provide the analytical starting point for these investigations.

\bibliography{references}

@article{ramesh2007multiferroics,
  title={Multiferroics: progress and prospects in thin films},
  author={Ramesh, Ramaroorthy and Spaldin, Nicola A},
  journal={Nature materials},
  volume={6},
  number={1},
  pages={21--29},
  year={2007},
  publisher={Nature Publishing Group UK London}
}

@article{wang2009multiferroicity,
  title={Multiferroicity: the coupling between magnetic and polarization orders},
  author={Wang, KF and Liu, J-M and Ren, ZF},
  journal={Advances in Physics},
  volume={58},
  number={4},
  pages={321--448},
  year={2009},
  publisher={Taylor \& Francis}
}

@article{fiebig2016evolution,
  title={The evolution of multiferroics},
  author={Fiebig, Manfred and Lottermoser, Thomas and Meier, Dennis and Trassin, Morgan},
  journal={Nature Reviews Materials},
  volume={1},
  number={8},
  pages={16046},
  year={2016},
  publisher={Nature Publishing Group}
}

@article{cheong2007multiferroics,
  title={Multiferroics: a magnetic twist for ferroelectricity},
  author={Cheong, Sang-Wook and Mostovoy, Maxim},
  journal={Nature materials},
  volume={6},
  number={1},
  pages={13--20},
  year={2007},
  publisher={Nature Publishing Group UK London}
}

@book{jackson1999classical,
  author    = {Jackson, John David},
  title     = {Classical Electrodynamics},
  edition   = {3},
  publisher = {John Wiley \& Sons},
  address   = {New York},
  year      = {1999},
  isbn      = {978-0-471-30932-1}
}

@book{landau1984electrodynamics,
  author    = {Landau, L. D. and Lifshitz, E. M.},
  title     = {Electrodynamics of Continuous Media},
  series    = {Course of Theoretical Physics},
  volume    = {8},
  edition   = {2},
  publisher = {Pergamon Press},
  address   = {Oxford},
  year      = {1984}
}

@article{bibes2008towards,
  title={Towards a magnetoelectric memory},
  author={Bibes, Manuel and Barth{\'e}l{\'e}my, Agn{\`e}s},
  journal={Nature materials},
  volume={7},
  number={6},
  pages={425--426},
  year={2008},
  publisher={Nature Publishing Group UK London}
}

@article{spaldin2005renaissance,
  title={The renaissance of magnetoelectric multiferroics},
  author={Spaldin, Nicola A and Fiebig, Manfred},
  journal={Science},
  volume={309},
  number={5733},
  pages={391--392},
  year={2005},
  publisher={American Association for the Advancement of Science}
}

@article{PhysRevB.92.134424,
  title = {Electromagnetically controlled multiferroic thermal diode},
  author = {Chotorlishvili, L. and Etesami, S. R. and Berakdar, J. and Khomeriki, R. and Ren, Jie},
  journal = {Phys. Rev. B},
  volume = {92},
  issue = {13},
  pages = {134424},
  numpages = {7},
  year = {2015},
  month = {Oct},
  publisher = {American Physical Society},
  doi = {10.1103/PhysRevB.92.134424},
  url = {https://link.aps.org/doi/10.1103/PhysRevB.92.134424}
}

@article{PhysRevLett.96.067601,
  title = {Ferroelectricity in Spiral Magnets},
  author = {Mostovoy, Maxim},
  journal = {Phys. Rev. Lett.},
  volume = {96},
  issue = {6},
  pages = {067601},
  numpages = {4},
  year = {2006},
  month = {Feb},
  publisher = {American Physical Society},
  doi = {10.1103/PhysRevLett.96.067601},
  url = {https://link.aps.org/doi/10.1103/PhysRevLett.96.067601}
}

@article{PhysRevB.91.041408,
  title = {Creation and amplification of electromagnon solitons by electric field in nanostructured multiferroics},
  author = {Khomeriki, R. and Chotorlishvili, L. and Malomed, B. A. and Berakdar, J.},
  journal = {Phys. Rev. B},
  volume = {91},
  issue = {4},
  pages = {041408(R)},
  numpages = {5},
  year = {2015},
  month = {Jan},
  publisher = {American Physical Society},
  doi = {10.1103/PhysRevB.91.041408},
  url = {https://link.aps.org/doi/10.1103/PhysRevB.91.041408}
}

@article{PhysRevB.94.104410,
  title = {Thermally induced magnonic spin current, thermomagnonic torques, and domain-wall dynamics in the presence of Dzyaloshinskii-Moriya interaction},
  author = {Wang, X.-G. and Chotorlishvili, L. and Guo, G.-H. and Sukhov, A. and Dugaev, V. and Barna{\ifmmode \acute{s}\else \'{s}\fi{}}, J. and Berakdar, J.},
  journal = {Phys. Rev. B},
  volume = {94},
  issue = {10},
  pages = {104410},
  numpages = {11},
  year = {2016},
  month = {Sep},
  publisher = {American Physical Society},
  doi = {10.1103/PhysRevB.94.104410},
  url = {https://link.aps.org/doi/10.1103/PhysRevB.94.104410}
}

@article{wanic2024magnetoelectric,
  title={Magnetoelectric fractals, Magnetoelectric parametric resonance and Hopf bifurcation},
  author={Wanic, M and Toklikishvili, Z and Mishra, SK and Trybus, M and Chotorlishvili, L},
  journal={Physica D: Nonlinear Phenomena},
  volume={467},
  pages={134257},
  year={2024},
  publisher={Elsevier}
}

@book{wiggins2003introduction,
  title={Introduction to applied nonlinear dynamical systems and chaos},
  author={Wiggins, Stephen},
  year={2003},
  publisher={Springer}
}

@article{arnold63,
  title={Proof of a theorem of AN Kolmogorov on the preservation of conditionally periodic motions under a small perturbation of the hamiltonian, Uspehi Mat. Nauk 18 (1963), no. 5, 13--40},
  author={Arnold, VI},
  journal={English translation: Russian Math. Surveys},
  volume={18},
  pages={9--36},
  year={1963}
}

@book{broer1996quasi,
  title={Quasi-periodic motions in families of dynamical systems: order amidst chaos},
  author={Broer, Hendrik W and Huitema, George B and Sevryuk, Mikhail B},
  year={1996},
  publisher={Springer}
}

@article{chotorlishvili2010quantum,
  title={Quantum chaos and its kinetic stage of evolution},
  author={Chotorlishvili, Levan and Ugulava, A},
  journal={Physica D: Nonlinear Phenomena},
  volume={239},
  number={3-4},
  pages={103--122},
  year={2010},
  publisher={Elsevier}
}

@article{PhysRevE.68.026216,
  title = {Overlapping of nonlinear resonances and the problem of quantum chaos},
  author = {Ugulava, A. and Chotorlishvili, L. and Nickoladze, K.},
  journal = {Phys. Rev. E},
  volume = {68},
  issue = {2},
  pages = {026216},
  numpages = {8},
  year = {2003},
  month = {Aug},
  publisher = {American Physical Society},
  doi = {10.1103/PhysRevE.68.026216},
  url = {https://link.aps.org/doi/10.1103/PhysRevE.68.026216}
}

@article{PhysRevE.70.026219,
  title = {Quantum-mechanical research on nonlinear resonance and the problem of quantum chaos},
  author = {Ugulava, A. and Chotorlishvili, L. and Nickoladze, K.},
  journal = {Phys. Rev. E},
  volume = {70},
  issue = {2},
  pages = {026219},
  numpages = {15},
  year = {2004},
  month = {Aug},
  publisher = {American Physical Society},
  doi = {10.1103/PhysRevE.70.026219},
  url = {https://link.aps.org/doi/10.1103/PhysRevE.70.026219}
}

@article{PhysRevE.71.056211,
  title = {Irreversible evolution of quantum chaos},
  author = {Ugulava, A. and Chotorlishvili, L. and Nickoladze, K.},
  journal = {Phys. Rev. E},
  volume = {71},
  issue = {5},
  pages = {056211},
  numpages = {9},
  year = {2005},
  month = {May},
  publisher = {American Physical Society},
  doi = {10.1103/PhysRevE.71.056211},
  url = {https://link.aps.org/doi/10.1103/PhysRevE.71.056211}
}

@article{Arnold1964,
  author  = {Arnold, V. I.},
  title   = {Instability of dynamical systems with several degrees of freedom},
  journal = {Soviet Mathematics Doklady},
  volume  = {5},
  pages   = {581--585},
  year    = {1964}
}

@article{Chirikov1979,
  author  = {Chirikov, B. V.},
  title   = {A universal instability of many-dimensional oscillator systems},
  journal = {Physics Reports},
  volume  = {52},
  number  = {5},
  pages   = {263--379},
  year    = {1979},
  doi     = {10.1016/0370-1573(79)90023-1}
}

@article{Nekhoroshev1977,
  author  = {Nekhoroshev, N. N.},
  title   = {An exponential estimate of the time of stability of nearly integrable Hamiltonian systems},
  journal = {Russian Mathematical Surveys},
  volume  = {32},
  number  = {6},
  pages   = {1--65},
  year    = {1977},
  doi     = {10.1070/RM1977v032n06ABEH003859}
}

@book{LichtenbergLieberman1992,
  author    = {Lichtenberg, A. J. and Lieberman, M. A.},
  title     = {Regular and Chaotic Dynamics},
  edition   = {2},
  series    = {Applied Mathematical Sciences},
  volume    = {38},
  publisher = {Springer-Verlag},
  address   = {New York},
  year      = {1992},
  doi       = {10.1007/978-1-4757-2184-3}
}

@article{DelshamsLlaveSeara2006,
  author  = {Delshams, A. and de la Llave, R. and Seara, T. M.},
  title   = {A geometric mechanism for diffusion in {Hamiltonian} systems
             overcoming the large gap problem: heuristics and rigorous
             verification on a model},
  journal = {Memoirs of the American Mathematical Society},
  volume  = {179},
  number  = {844},
  year    = {2006},
  doi     = {10.1090/memo/0844}
}

@Article{Maroulakos2026,
  author    = {Maroulakos, D. and Wal, A. and Tralle, I. and Mishra, Sunil K. and Chotorlishvili, L.},
  journal   = {Nonlinear Dynamics},
  title     = {Hopf bifurcation and stochastic spiking in an antiferromagnetic {FitzHugh-Nagumo} normal form},
  year      = {2026},
  issn      = {1573-269X},
  month     = Oct,
  number    = {20},
  volume    = {114},
  doi       = {10.1007/s11071-026-13102-6},
  publisher = {Springer Science and Business Media LLC},
}

\end{document}